\documentclass[]{aa}
\usepackage{graphicx}
\usepackage{txfonts}
\usepackage{subcaption}
\usepackage{natbib,twoopt}
\usepackage[breaklinks=true]{hyperref} 
\bibpunct{(}{)}{;}{a}{}{,} 
\makeatletter
\newcommandtwoopt{\citeads}[3][][]{\href{http://adsabs.harvard.edu/abs/#3}%
{\def\hyper@linkstart##1##2{}%
\let\hyper@linkend\@empty\citealp[#1][#2]{#3}}}
\newcommandtwoopt{\citepads}[3][][]{\href{http://adsabs.harvard.edu/abs/#3}%
{\def\hyper@linkstart##1##2{}%
\let\hyper@linkend\@empty\citep[#1][#2]{#3}}}
\newcommandtwoopt{\citetads}[3][][]{\href{http://adsabs.harvard.edu/abs/#3}%
{\def\hyper@linkstart##1##2{}%
\let\hyper@linkend\@empty\citet[#1][#2]{#3}}}
\newcommandtwoopt{\citeyearads}[3][][]%
{\href{http://adsabs.harvard.edu/abs/#3}
{\def\hyper@linkstart##1##2{}%
\let\hyper@linkend\@empty\citeyear[#1][#2]{#3}}}
\makeatother

\usepackage{color}

\newcommand{\bifrost}{{\textit{Bifrost}\ }}

\newcommand{\angstrom}{\text{\normalfont\AA}}

\begin{document}

\title{Investigating 4D Coronal Heating Events In MHD Simulations }

\author{Charalambos Kanella\inst{\ref{affil1},\ref{affil2}}
          \and
          Boris V. Gudiksen\inst{\ref{affil1},\ref{affil2}}}

\institute{Rosseland Centre for Solar Physics, University of Oslo, P.O. Box 1029 Blindern, NO-0315 Oslo, Norway\label{affil1}
\and 
Institute of Theoretical Astrophysics, University of Oslo, P.O. Box 1029 Blindern, NO-0315 Oslo, Norway\label{affil2} \\
\email{charalambos.kanella@astro.uio.no}}

\abstract
{One candidate-model for heating the solar corona is magnetic reconnection that embodies Ohmic dissipation of current sheets. When numerous small-scaled magnetic reconnection events (in observations  are the speculated \emph{nanoflares}) occur, then it is possible to heat the corona.}   
{Due to the limitations of  current instrumentation, nanoflares cannot be resolved. But their importance is evaluated via statistics by finding the power-law index of the energy distribution. This method is however biased due to technical and physical reasons. We aim to overcome limitations imposed by observations and statistical analysis. This way, we will identify,  and study the small scale impulsive events.}
{We employ a three-dimensional magnetohydrodynamic (3D-MHD) simulation using the \bifrost code. We also employ a new technique to identify the evolution of 3D Joule heating events in the corona.  Then, we derive parameters describing the heating events in these locations, study their geometrical properties and where they occur with respect to the magnetic field.} 
{We report on the identification of heating events. We obtain the distribution of duration, released energy, and volume. We also find weak power-law  correlation between these parameters.  In addition, we  extract information about geometrical parameters  of 2D slices of 3D events, and  about the evolution of resolved Joule heating compared to the total Joule heating and the magnetic energy in the corona. Furthermore, we identify relations between the location of heating events and the magnetic field. }
{Even though the energy power index is less than 2, when classifying the energy release into three categories regarding with respect to the energy release (pico-, nano-, and micro-events), we find that nano-events release $82 \ \%$ of the resolved energy. This fraction corresponds to an energy flux larger than the one needed to heat the corona. Though, no direct conclusions can be draw, it seems that the most popular population among small-scale events is the one that contains short-lived, nano-scaled energetically events with small spatial extend. Generally, the locations and size of heating events are affected by the magnetic field magnitude.}

\maketitle

\section{Introduction}\label{sec:intro}

The  mechanical energy contained in the flows of the convective zone and the photosphere is so big that only a fraction is needed to heat the solar corona \citep{Gesztelyi1986}. It is conventional to attribute the medium of transferring the energy generated by the mechanical drivers to the magnetic field.  Other mechanisms have been proven  not to work. For instance, energy cannot be transported from the photosphere to the corona via mass convection, neither via sound waves because this class of waves is dissipated, or reflected before reaching the corona \citep{Carlsson2002,Carlsson2007}.  Small velocity amplitude MHD waves can reach the corona, but those do not carry enough energy  \citep{Hara1999, Tomczyk2007}. The only waves that can penetrate into the region, and transport enough energy, are Alfv\'en  waves, however dissipating them is not so easy \citep{vanBallegooijen2011, AsgariTarghi2012}.

The magnetic field, anchored in the photosphere, extends throughout the solar atmosphere, establishing a link between the photosphere and the corona. This link enables mechanical energy to propagate towards the corona via Poynting flux  \citep{Klimchuk2006,Hansteen2015}. There are two components of the vertical Poynting flux, the horizontal motions of the vertical component of magnetic field and the transport of the horizontal field by vertical motions. Both of these components transport energy into the corona.

The energy carried by the Poynting flux is stored in the form of currents. The process involves injection of energy  that transforms a potential into a non-potential field. As a consequence, magnetic field gradients appear, which  are responsible for current sheet formation.  Current sheets store any excess energy above the energy level of a potential field \citep{Galsgaard1996, Gudiksen2005}. In MHD, current sheets express the distortion of the magnetic field, i.e., $\textbf{J}\propto \nabla \times \textbf{B}$. However, the magnetic field cannot store infinite energy. At some point, a critical value is reached, and energy is released impulsively in a stochastic manner. The maximum amount of energy that can be released is the non-potential magnetic energy, which is replaced continually due to the motions of the mechanical drivers in photosphere and the convective zone.  As it was shown by \cite{Hansteen2015}, the total energy input in the coronal region is ``spatially intermittent and temporally episodic'', but in the long-term heating is almost constant.

The inclination between currents in current sheets and a magnetic field plays an important role on the work done. When a current is aligned with the magnetic field, then the exerted Lorentz force on the plasma is zero,  i.e., $\textbf{J} \times \textbf{B}=0$. However, when there is an inclination between current and magnetic field,  the Lorentz force is then non-zero and work is done.  Then, energy stored in the currents in the presence of finite resistivity is dissipated, and cross-field currents release a part of the stored energy via Joule heating \citep{Low1990}. If currents (and thus electric field) are perpendicular  to the magnetic field, then  magnetic field topology changes significantly and magnetic reconnection occurs \citep{Parker1972}.

The non-potential magnetic field can be mapped through Quasi-Separatrix-Layers (QSL) \citep{Aulanier2006}; QSLs are the equivalent to separatrices in 2D. While the stressing of the magnetic field continues and currents form, QSLs become thinner and magnetic field distortions larger until reconnection takes place \citep{Aulanier2006}. The high current density, helps to increase the resistivity locally and allows magnetic flux of opposite polarity to reconnect. During magnetic reconnection several processes take place, such as direct thermal heating via Joule heating, energy transport via acceleration of particles, excitation of waves and shock generation.

Current sheets have  scales that vary in a hierarchical manner from bigger to smaller scales. Current sheets can reach scales so small that magnetic energy dissipation via Joule or viscous heating is feasible. Fragmentation of current sheets occurs mostly in regions with very large resistivity. The fragmentation stops when currents have scales, where resistive diffusion \citep{Nordlund2012}, or friction can act \citep{VanBallegooijen1986}.  Currents evolve on  similar timescales as the magnetic field does. According to \cite{Galsgaard1996}, current sheet formation takes a few seconds, while current sheet dissipation can take from few to thousands of seconds.

The observational traces of magnetic reconnection are flares.  Flares range in energy output from large ($10^{32}$ erg) to the smallest postulated but so far unobserved nanoflares ($10^{24}$ erg), spanning many orders of magnitude. 


Statistics of flares are important because nanoflares, according to \citet{Parker1972,Parker1988}, can heat the solar corona, if a very large number of them occurs.  According to observations, the frequency of energy release from flaring events is distributed as a power-law function $N(E) \propto E^{\alpha_E}$, where $\alpha_E$ is the power index of energy, and $N(E)$  the number of events  in the energy range $E$ and $E + \delta E$. If the index is larger than  two, then nanoflares are more important energetically than large flares  \citep{Hudson1991}. Constraining the value of the powerlaw index has been the goal of numerous observational campaigns and investigations, but the value of the of the power law index is still disputed. Examples of such observations are the following:  in peak of HXR, \cite{Christe2008} have found  $1.58\pm0.02$;  in fluence of HXR, \cite{Enriquez1999} and \cite{Crosby1993} have  found  $1.39\pm0.01$, and $1.48\pm0.02$ respectively;  in fluence of SXR and  peak of SXR, \cite{Drake1971} have found $1.44$ and $1.75$ respectively; in peak of UV and EUV intensities, \cite{Aschwanden2002}  have found $1.71\pm0.1$ in  $171 \angstrom$, $1.75\pm0.07$ in  $191 \angstrom$, and $1.52\pm0.1$ in Aluminium-Magnesium filter on Yohkoh spacecraft. Since very small events cannot be resolved by the current instrumentation, and the observed power-law indices are smaller than two, taking the raw numbers from these works indicates that the powerlaw index is less than two,  suggesting that large flaring events are more significant energetically than smaller ones.

Other quantities that describe heating events also follow power-laws. For example, the duration of each event exhibits a power-law slope in observations that depends on the solar cycle. The slope  has minimum value during the solar minimum, and maximum during solar maximum. In fact, \citet{Aschwanden2012} found in 35 years of GOES observations that during solar minimum the slope is as small as 2, while during solar maxima the slope ranges from 2 to 5. In the literature, the volumes of flares are usually calculated by making string assumptions, making them less reliable, but producing a power-law distribution with power-law indices that varying between 1.5 and 2.08 (examples are in table 9 in \citet{Aschwanden2014}).

However, finding the power-law index for flare distributions is not trivial due to observational biases.  Finding the volume of a flare is difficult due to our inherently 2D observations. Both background and foreground contamination makes the estimation of the distance taken up by the flare along the line of sight very difficult. The determination of thermal energies requires the knowledge of the volume occupied by the flaring events \citep{Benz2002}. We are only able to deduce information from observations about the area perpendicular to the line of sight, and therefore scaling-laws depend on assumptions in order to calculate the volume (e.g. \cite{Benz2002,Shimizu2013}). There is no direct connection between the dimension of a flare in each of the three spatial dimensions, so we cannot find the volume of a flare from two measured dimensions \citep{Morales2009}. The passband used for the observations also produces different projected areas since they are sensitive to gas at different temperatures and that the densities at the different temperatures are rarely equal. Finding the duration of flares is not trivial, because  flare identification algorithms depend on the  the identification technique and the criteria used therein. These problems create uncertainties in the estimated parameters of the flares. 

Sampling or selection bias is another problem that is rarely taken into account. Typically, the method used to detect and select flaring events produce these biases. The synchronicity of observations from different passbands has different effects on small and large flares. Short events are affected by the integration time, either because the events are drowned out by background (if the integration time is long) or under sampled (if the time between exposures are long). Resolution also  under-represents low energy events, because small events produces smaller peaks if they have sub-pixel sizes. The larger flares can be subject to biases if the total observation time is too short. Finally, distributions can be skewed if a large number of small unresolved events are labeled as a single large event. The fitting method, the error bars used in fitting, and the correct choice of background heating and noise subtraction affect the power-law index \citep{Benz2002}.  As stated also by  \cite{Hannah2011}, the large range of power-law slopes found in different studies from various researchers is also a product of the method used to extract results and the instrumentation employed in different periods during the solar cycle.


{In this study, our most important goal is to study three-dimesnional heating events related to magnetic reconnection, and evaluate their contribution in heating the corona. We need to do that in an experiment that overcomes most of the observational restrictions. To achieve that, we simulate the solar atmosphere using the \bifrost code \citep{Gudiksen2011}, and use a relatively new method to identify three-dimensional heating events, and we followed their evolution in time.


Being able to identify 3D events gives us the opportunity to study them in detail. More specifically, we want to check to what extend small-scale events contribute to coronal heating, and to identify if there is a lower energy cut-off.  In addition, we want to asses the contribution of Joule heating events with respect to the total Joule heating and the magnetic energy in the corona. We also want to explore how heating events manifest themselves in 3D space, and check their evolution in time. Another objective of this study is to check if we can identify any scaling-laws between energy, duration, and volume that could help observers to derive conclusions by observing one instead of another. Moreover, we want to locate where heating events occur with respect to the magnetic field and compare the results with the literature.


This paper discusses the properties of  heating events related to magnetic reconnection that have been identified in a 3D simulation. We study their individual and collective behaviours under the prism of coronal heating. The remainder of this paper is organised in the following way: in Sect. \ref{subsec: Bifrost}, we briefly describe the \bifrost code \citep{Gudiksen2011}; in Sect. \ref{subsec:Identification_method}, we describe the method used to identify the evolution of Joule heating events, and the rest of the parameters. Section \ref{sec:Results} reports on the findings. More specifically, Sect. \ref{subsec:geom_asp}  includes the results of our investigation on the geometrical properties of the 3D structures we identify, while Sect. \ref{subsec:Hists} contains the distributions and power-law fits of duration and  energy together with the cumulative distribution function of the mean volume. A statistical analysis of several parameters is in Sect.  \ref{subsec:statistical_analysis}. Finally, in Sect. \ref{sec:conclusions}, we discuss our findings and derive  conclusions.

\section{Method} \label{sec:Method}

In the current section, we briefly describe the \bifrost code used to create the snapshots of the solar atmosphere we will be analysing in this work. We also describe the method employed to detect heating events spatially and temporally in the region of interest. 

\subsection{\bifrost simulation} \label{subsec: Bifrost}
The \bifrost code \citep{Gudiksen2005, Gudiksen2011} is a 3D MHD code that can simulate a stellar atmosphere from the convective zone up to the corona. It can include numerous special physics and boundary conditions to adequately model stellar atmospheres. It solves a closed set of MHD partial differential equations along with equations describing radiation transport, thermal conduction along the magnetic field and a realistic equation of state. A  Cartesian grid is used to solve  the system of equations  using 6th order differential operators, 5th order interpolation operators and a 3rd order Hyman method with variable time-step. The  description of the non-grey radiative transfer includes the scattering between optically thin and thick regions of the photosphere and chromosphere to properly model the region \citep{Hayek2010}, and a chromospheric radiation approximation where the energy balance is critically dependent on the scattering in strong spectral lines, and optically thin radiation in the upper atmosphere.

The energy equation used in \bifrost is of special interest in this work. The radiative and conductive processes  can be described though the following equation of the evolution of internal the energy: 
\begin{equation} \label{eq:energy}
\frac{\partial e}{\partial t} +  \vec{\nabla} \cdot e\vec{u} = Q_{\text{C}} + Q_{\text{R}} - P \vec{\nabla} \cdot \vec{u} + Q_{ \text{J}} + Q_{ \text{Vi}} 
\end{equation}

where $e$ is the internal energy per unit volume, $\vec{u}$ the velocity vector, $P$ the gas pressure, $Q_{\text{C}}$ the contribution from the Spitzer thermal conduction along the magnetic field \citep{Spitzerbook}. $ Q_{ \text{J}}$ represents the Joule heating, $Q_{ \text{Vi}}$ is the viscous heating and $Q_{\text{R}}$ the energy contribution from the emitted or absorbed radiation. 

In this paper, we also employ  the simulation used by \cite{Kanella2017a}. We use the data from a simulation that includes a region enclosed between the solar convective zone and the corona. The simulated convective zone extends 2.5 Mm below the photosphere, and the simulated box reaches 14.3 Mm above the photosphere. In the vertical direction z, the upper boundary is open, while the lower boundary maintains the convective flow by giving the inflowing gas enough entropy to maintain the correct effective temperature of the solar photosphere, i.e., 5780 K. In the horizontal x-y plane, the numerical volume is periodic.  

The simulation box contains $768 \times 768 \times 768$ cells and spans a physical volume of $24 \times 24 \times 16.8 $ Mm$^3$.  The horizontal grid spacing ($dx = dy$) is constant and equal to 31.25 km, while the vertical grid spacing varies to resolve the magnetic field, temperature and pressure scale heights. The vertical spacing ($dz$) is roughly 26 km in the photosphere and the chromosphere, and increase slowly up to 165 km at the upper boundary. This simulation was created to resemble  a structure of  magnetic field network embedded in the quiet Sun (QS). The configuration contains two relatively strong magnetic regions of opposite polarity, which are connected with a magnetic structure that has a loop-like shape. Throughout the simulation a horizontal field of 100 Gauss is injected in the inflowing regions at the lower boundary. This injection maintains the well-known salt and pepper magnetic field. More detailed description of the simulation setup can be found in \citet{Carlsson2016}; the only difference is that the one described in \citet{Carlsson2016} also incorporates the effects of non-equilibrium ionization of hydrogen. 

For our analysis, we chose a region of interest (ROI) that corresponds to the corona. The ROI starts at height  3.28 Mm above the photosphere, where the temperature is equal to 1 MK, and extends up to the top of the simulation box excluding a few cells zones because they are affected by boundary conditions. Therefore, the volume of interest is $24 \times 24 \times 9.5$ Mm$^3$, and corresponds to $768 \times 768 \times 331$ grid cells.

\subsection{Identification method}\label{subsec:Identification_method}

 
To quantitatively study the effects of magnetic reconnection, we choose to analyse the Joule heating term in equation \ref{eq:energy}. The grid-size is of the order of a few decades of kilometres, and represents  scales that are much larger than the physical scales at which  physical resistivity and viscosity are effective. Therefore, \bifrost uses the minimum numerical diffusivity (resistivity), which ensures the stability of the code. Further details about the numerical resistivity and the heating term can be found in our previous work \citep{Kanella2017a} and in \citet{Gudiksen2011}. 

In our previous work \citep{Kanella2017a} we described the details of the numerical tool, i.e., ``ImageJ''  \citep{Collins2007,Ollion2013}, and algorithm  used to identify three-dimensional structures in  each snapshot, i.e., Adaptive Generic Iterative Thresholding Algorithm (AGITA) \citep{Ollion2013,Gul-Mohammed2014}. Here, we describe the identification method and the quantity used for this purpose in different terms so as to understand the underlying process.

Magnetic reconnection is a topological phenomenon, therefore the identification of each event is difficult without a detailed study of each region, however the implicit effect of reconnection could be located. The best proxy we have to study such topological events is to investigate the Joule heating term in the \bifrost code. Joule heating depends, among other, on current density, which in MHD express the degree of magnetic field distortion, or the magnetic field gradients.  When there is inclination between the magnetic field and currents, then work is done, and it is a requirement that part of the current is perpendicular to the magnetic field in order for magnetic field to reconnect.

The method we employ here and in \citet{Kanella2017a}  depends on the ability of the algorithm to find spikes (local maxima) in the Joule heating in the three-dimensional space, and follow the negative gradient in all directions until the gradient at some level of the heating, i.e., $E_0$, becomes zero. The 3D iso-surface of the Joule heating at level $E_0$ around that local maxima gives the event volume, and thus the total energy of the event can be calculated. The method is then repeated for the next local maxima. The strongest point of the method  is the fact that we can use different values for $E_0$ for each spike, save those results and choose which feature we consider the best option. For this purpose, we use pre-specified geometrical criteria, such as the largest volume between pre-specified limits. It is important to point out, that this method does not attribute all the Joule heating to identified heating events. A significant amount of Joule heating is not attributed to events, either as a consequence of the choice to stop the event volume at the $E_0$ level or simply  because the events are not strong or large enough \citep[see][]{Kanella2017a}.

We perform the same procedure for 57 simulation snapshots, which are separated by 10 seconds, starting from $t_1=830$ s of  solar time in our simulation.

In order to find the connection between features, and establish the link between four-dimensional structures, we follow the evolution of each feature.  Starting from snapshot $i$, at time $t_i$, we check if there are other feature(s) at the same coordinates in the next snapshot $t_i+10 \ $s, if yes then they share a common label, and our algorithm checks in the new coordinates of the new features (at $t_i+10$ s) for any  features in the next snapshot (at $t_i+20 $ s). The procedure continues to the next snapshot until no feature is identified, then the algorithm proceeds to the next feature in snapshot $i$ at time $t_i$. Then, the same procedure continues for the features of the next snapshot, but only for those that  had not been connected with other features in prior steps. Summarising the process, features that overlap even with one pixel in the 4th dimension are considered to be one single event progressing in time.

\section{Results} \label{sec:Results}

Using our method, we identify 145306 features in  570 s of solar time. Figures \ref{fig:qj_vs_magnetic_field} and \ref{fig:labels_vs_magnetic_field} illustrate examples of our findings at $t=1130$ s. We plot field lines over Joule heating events to  render the magnetic field topology in the corona with respect to the location of events. We choose 75 by 75 starting points for the field lines, equally distributed in the horizontal plane at the base of the corona. In  Fig. \ref{fig:qj_vs_magnetic_field}, we represent the 2D slices of 3D Joule heating events  at the base of the corona together with contours of the vertical component of the magnetic field. Our  aim is to identify  any possible correlation between magnetic field and density of heating events.

\begin{figure}
\centering
\includegraphics[width=\hsize]{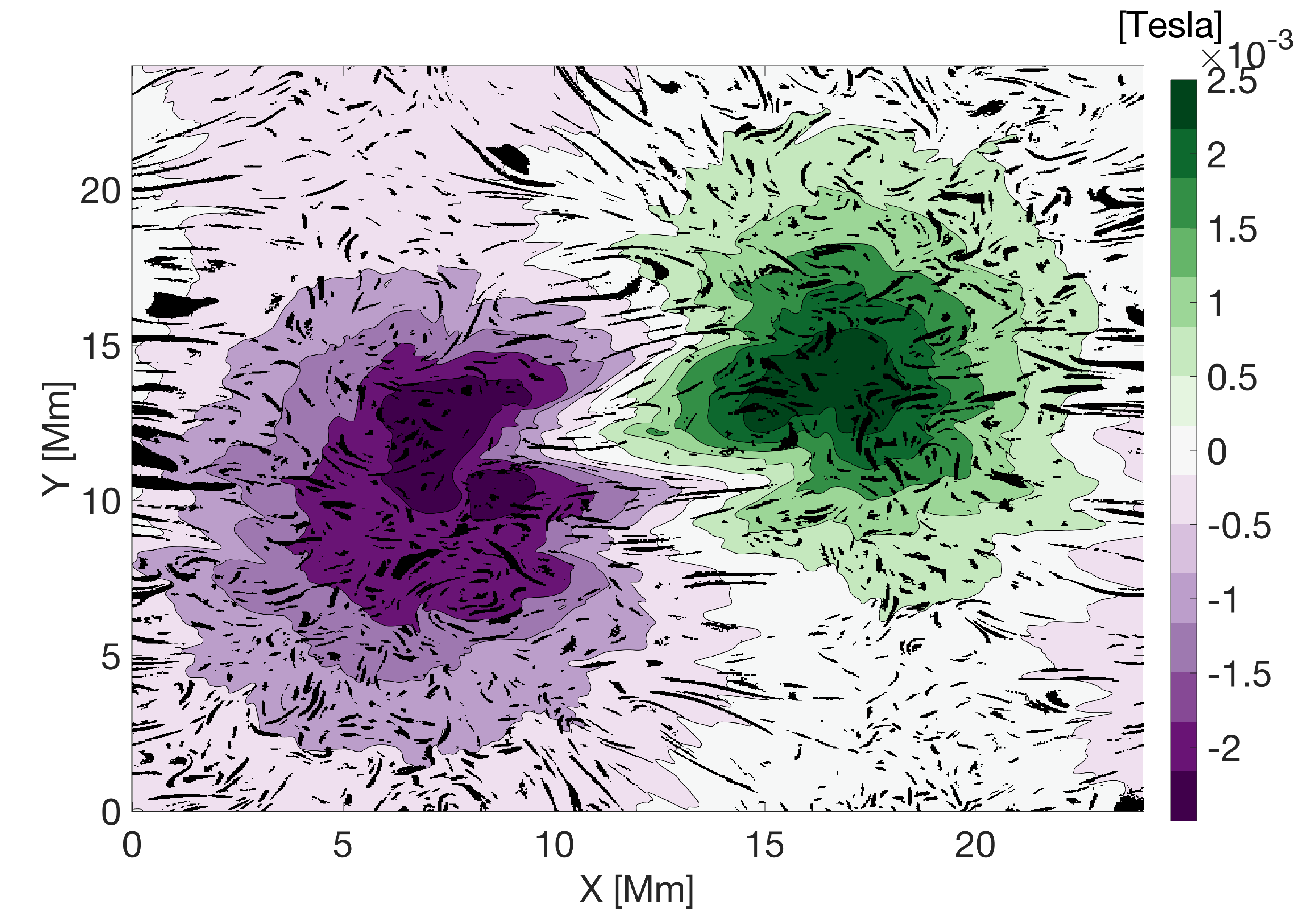}
\caption{Contours of the vertical component of the magnetic field at the base of the corona (at $t=1130$ s) together with the bases of the identified Joule heating events. \label{fig:qj_vs_magnetic_field}}
\end{figure}

\begin{figure}
\begin{subfigure}{0.5\textwidth}
\centering
\includegraphics[width=\hsize]{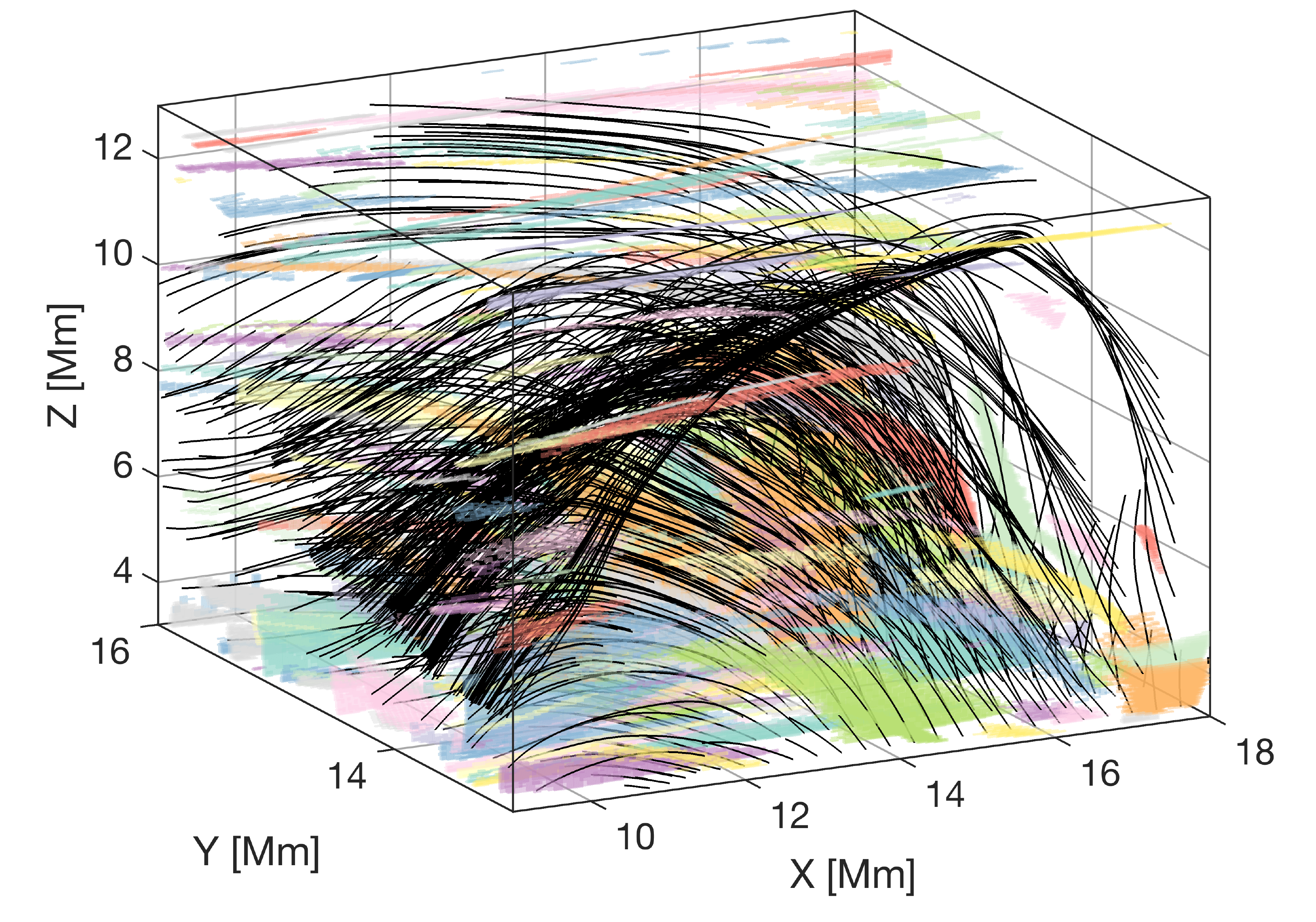}
\end{subfigure}
\begin{subfigure}{0.5\textwidth}
\centering
\includegraphics[width=\hsize]{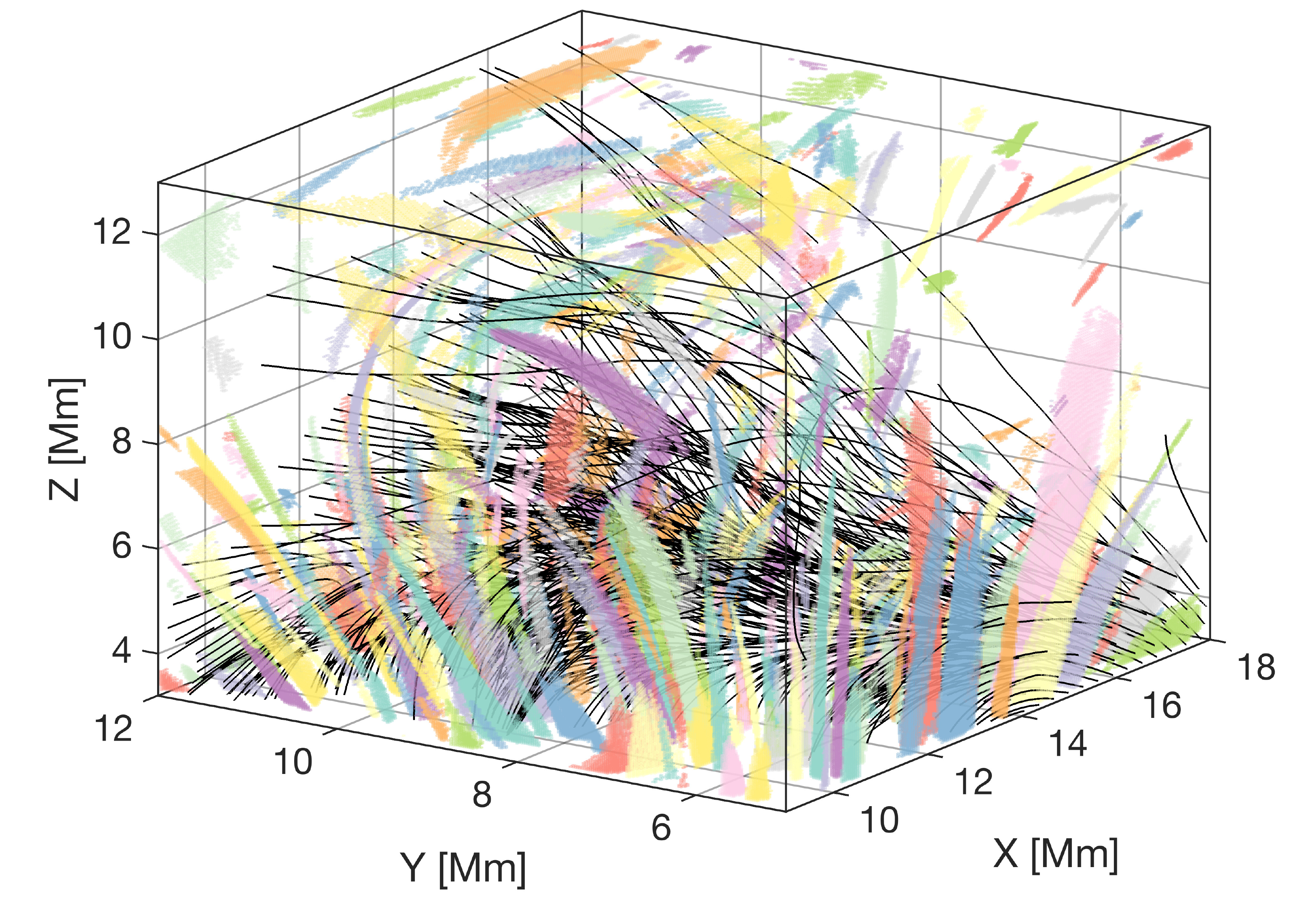}
\end{subfigure}
\caption{Magnetic field lines over-plotted together with identified features, in which each colour represents a different feature, at $t=1130$ s. Different panels illustrate different orientations of the x-y plane. \label{fig:labels_vs_magnetic_field}}
\end{figure}

We calculate the amount of Joule heating attributed to heating events by our algorithm in every snapshot. The evolution of the resolved energy density in the ROI is presented in Fig. \ref{fig:resolved_E_percentage}, along with the evolution of  the energy density of the magnetic field, and total energy density of the Joule heating. Note that we employed the identification method on the energy density rate, and we have converted the quantity to energy density by multiplying by the duration between snapshots. 
 
The difference between total Joule heating and Joule heating attributed to heating events is what we call here \emph{residual} heating.  We speculate that source of the residual energy is  a combination of background heating, numerical noise, and unresolved events. Background heating may be due to a lower energy release mechanism that heats the region in a less or non-impulsive manner. For example, MHD waves that could induce currents, or remnants of current sheets after an impulsive event that burn slowly  \cite{Janvier2014}. Another possibility is the equivalent of the original nano-flare picture by \citet{Parker1983b}, where all flares were collections of nano-flares, in small or large numbers, and using this method to identify events we suffer the same problems as observers do, that we cannot distinguish a sea of low energy events, from an almost constant background heating. 

\begin{figure}
\centering
\includegraphics[width=\hsize]{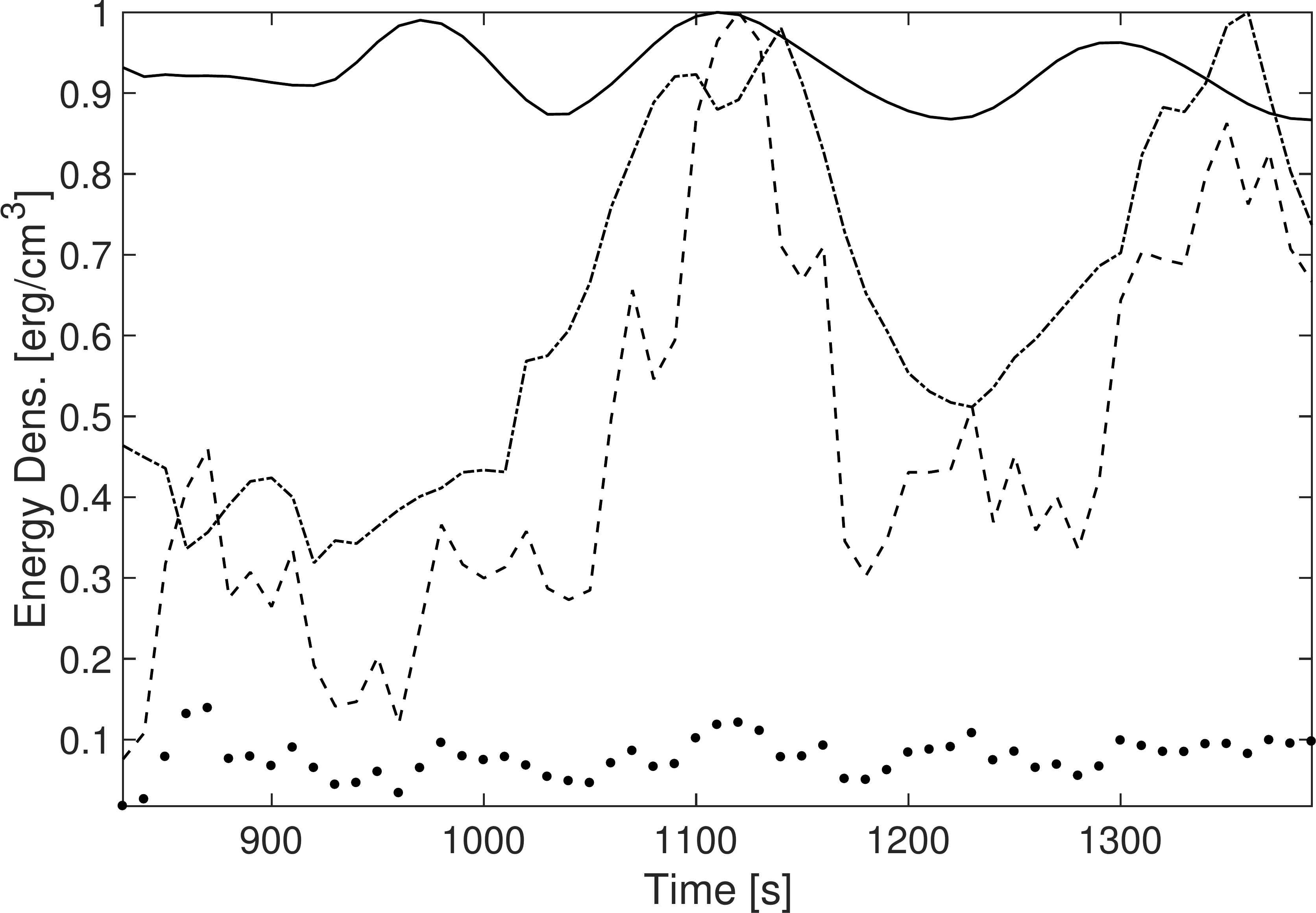}
\caption{Plot of  evolution of the following normalised quantities: Magnetic energy density (line) with maximum value $4.6 \ 10^{8} \ \textrm{erg}/\textrm{cm}^{3}$, Joule heating density of the identified features (dashed line) with maximum value $10^{5} \ \textrm{erg}/\textrm{cm}^{3}$, total Joule heating density released in the ROI (dashed dotted line) with maximum value $9.3 \ 10^{5} \ \textrm{erg}/\textrm{cm}^{3}$, and ratio between resolved and total Joule heating density (dots). \label{fig:resolved_E_percentage}}
\end{figure}

Assuming that the events we do identify are not just a conglomeration of much smaller events, an interesting aspect of simulating the Sun is that you can resolve 3D heating events and follow their evolution. In Fig. \ref{fig:power_vs_duration_subplot}, we show the evolution of energy release rate in four cases. Panels a and  b represent long duration energetic events, which release energy in a non-monotonic fashion; Panel c illustrates energy release of a nanoflare-like event, resolved by 4 steps.  Assuming that a heating event should have impulsive and gradual phases, panel d depicts an unresolved heating event because it has only a decline phase.

\begin{figure}
\centering
\includegraphics[width=\hsize]{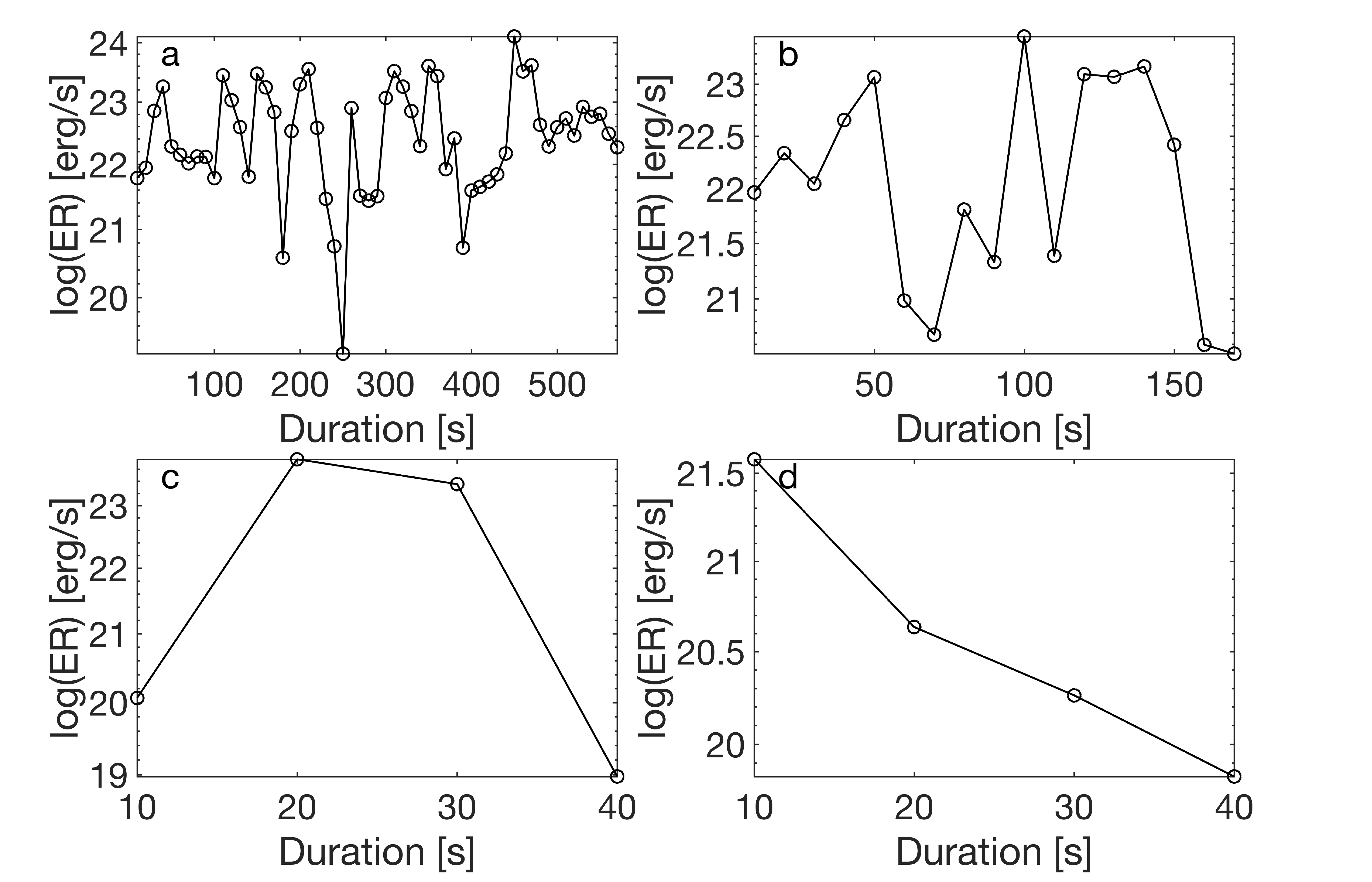}
\caption{Four cases of  evolution of energy rate in heating events. Energy rate (ER) of each event is presented with an opened cycle in logarithmic scale at each time-step; two successive points have an interval of 10 s.     \label{fig:power_vs_duration_subplot}}
\end{figure}


\subsection{Geometrical parameters }\label{subsec:geom_asp}

The shape and volume of each identified event varies significantly. To quantify that, we explore three geometrical parameters of 2D slices of a 3D event with respect to height z after we fitted an ellipsoid to each slice. We choose to fit an ellipsoid at each feature's 2D slice, because the majority of shapes in the  horizontal slice of the simulation box  at the base of the corona, as illustrated in Fig. \ref{fig:qj_vs_magnetic_field}, could be approximated with such surface. The parameters describing an ellipsoid are easy to be understood, and the process to do so is very easy and reproducible. The parameters we investigate are the following: cross-section (area), eccentricity, and orientation (between -90 to 90 degrees) of the ellipsoid's major axis with respect to the x-axis. In Fig. \ref{fig:geometrical_param}, we plot two examples of two apparently different shapes. 

We expect the area to  increase or decrease coherently until the limit of our conservative resolution, i.e., around 4500 $\textrm{km}^2$, unless a sudden and large magnetic field distortion occurs locally. In such case, the geometrical parameters could change irregularly. 

The example on the right of Fig.  \ref{fig:geometrical_param} represents a structure, which has a very thin upper half part close to the resolution limit, hence we observe sharp spikes in the changes in orientation and eccentricity along height which are probably not physical, but simply an effect of the resolution.



\begin{figure}
\begin{subfigure}{0.5\textwidth}
\centering
\includegraphics[width=\hsize]{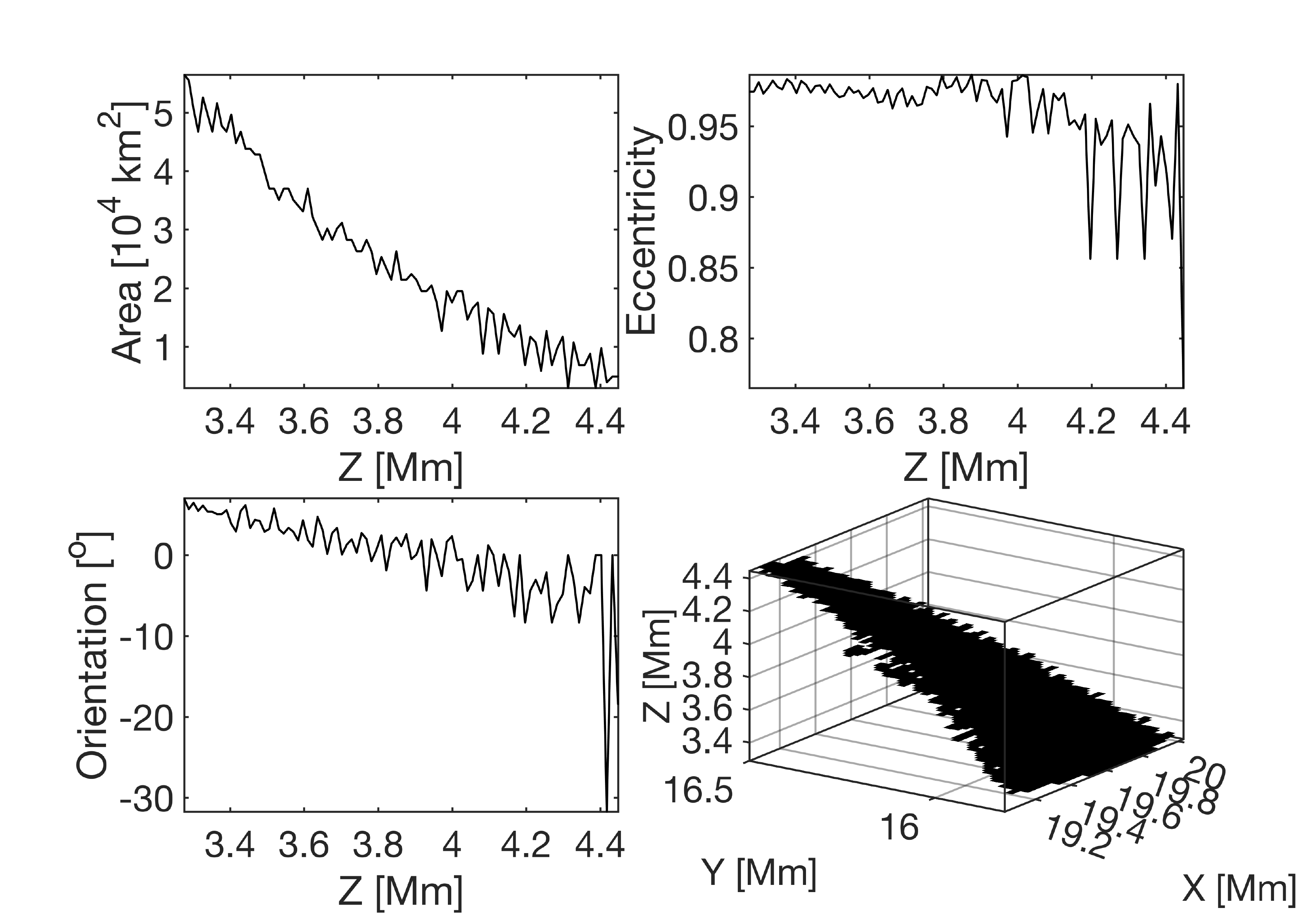}
\end{subfigure}
\begin{subfigure}{0.5\textwidth}
\centering
\includegraphics[width=\hsize]{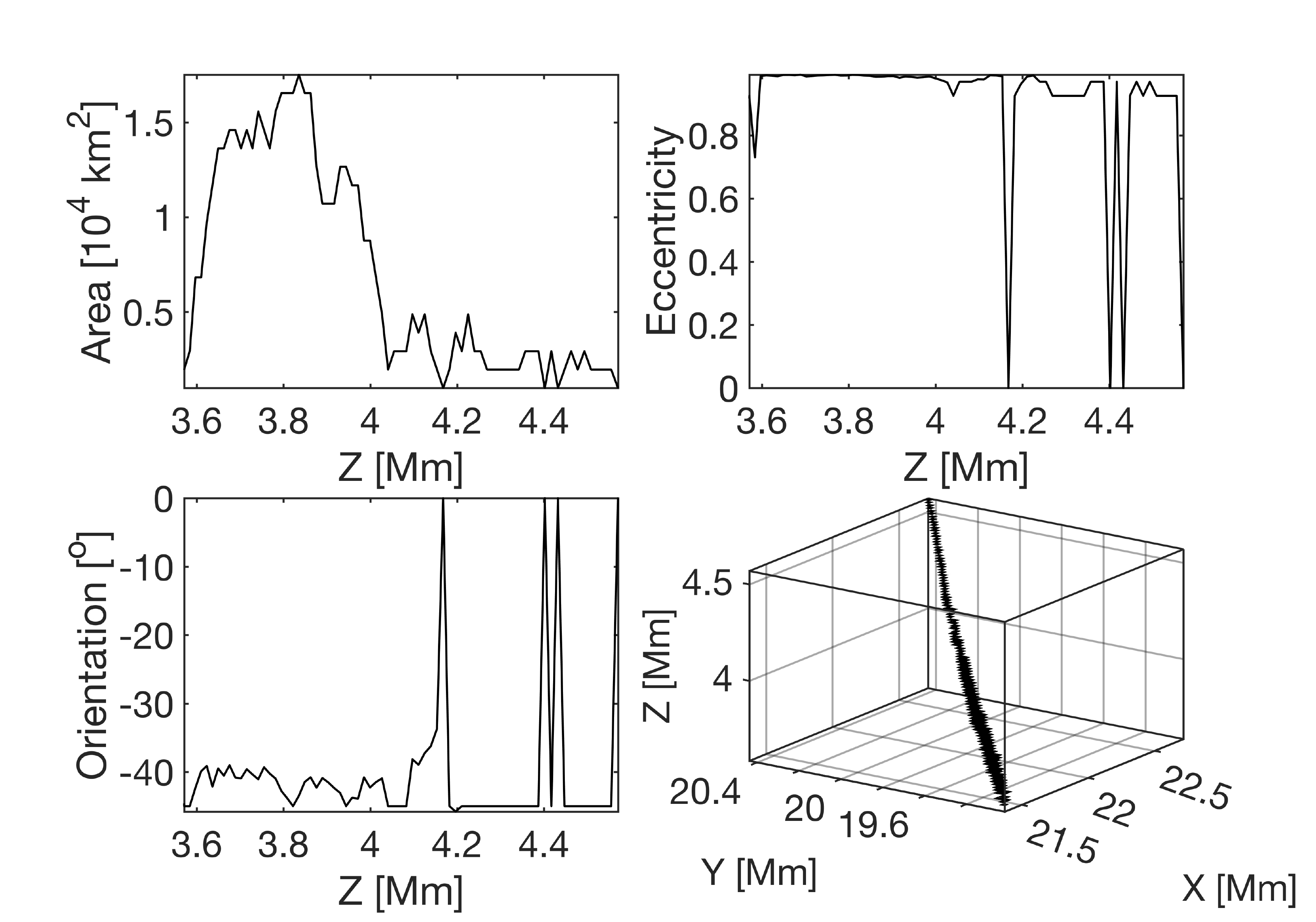}
\end{subfigure}
\caption{ Three geometrical parameters (area,  orientation, and eccentricity) of 2D slices of two different 3D events calculated along height. The examples are identified at $t=1130$ s of solar time and exhibit fun and spine-like shapes.  The 3D structure is also presented in each case.  \label{fig:geometrical_param}}
\end{figure}

\subsection{Histograms: energy, mean volume and duration}\label{subsec:Hists}

Isolating heating events enables us to explore different parameters, such as energy release,  mean volume, and duration of heating events. Due to the fact that the volume of each identified event evolves and changes with time, we calculate the mean volume of each identified event throughout its evolution. Mean volume is the total of volumes of an identified event at each snapshot for its total duration divided by the number of snapshots.  These parameters can be interpreted  collectively via histograms. For this reason, we calculate the differential size distribution (DSD), i.e., number of identified events per logarithmic bin-width. In cases where the DSDs  can be approximated by a power-law distribution, we fit one that has the following expression:  

\begin{equation} \label{eq:DSD}
\frac{dN(x)}{dx}=A \ x^{-\alpha}. 
\end{equation}

where the left hand side is the DSD, $\alpha$ the power index and $A$ a constant. 

The bin-width or the number of bins is chosen with the Freedman-Diaconis rule, which is not much sensitive to outliers, and it is suitable for data with heavy-tailed distributions. It uses a bin-width equal to $2 \times IQR(x)\times N^{-1/3}$, where $IQR$ is the interquartile range of the data, and $N$  is the number of observations in the sample $x$.

Energy and event duration exhibit power-law distribution as illustrated in Figs. \ref{fig:hist_energy} and \ref{fig:hist_duration} respectively.  To find the power index, we fit power-law functions using the $\chi^2$-minimisation technique. However, due to the knee on the lower end in the energy histogram, we choose the maximum DSD value and the corresponding parameter value to be the lower boundary at which we fit the power-law function. The minimum parameter value is considered to be the minimum resolved value and that is $E_0= 1.1 \ 10^{20}$ erg. The power-law index is  $\alpha=1.41 \pm 0.01$ and is fitted over $91 \ \%$ of total number of  events. The energy released by the events that are not included in the power-law fitting have insignificant contribution to corona heating. The fitted power-law in the duration histogram, uses the total number of identified events, and the slope is $\alpha=2.87 \ \pm 0.01$.

In the histogram of mean volume (i.e., Fig. \ref{fig:hist_mean_vol}), we find that data cannot be approximated by a power-law distribution, hence we find that the best way to describe the mean volume is via a cumulative distribution function (CDF). The mean volume spans three orders of magnitude from volumes around $10^{21} \  \textrm{cm}^3$ up to volumes around $10^{24} \ \textrm{cm}^3$. We find that, the CDF is very steep in the first $85 \ \%$  of volumes (volumes less than $2\ 10^{22} \ \textrm{cm}^3$), whereas the distribution in the rest becomes flatter.

\begin{figure}
\centering
\includegraphics[width=\hsize]{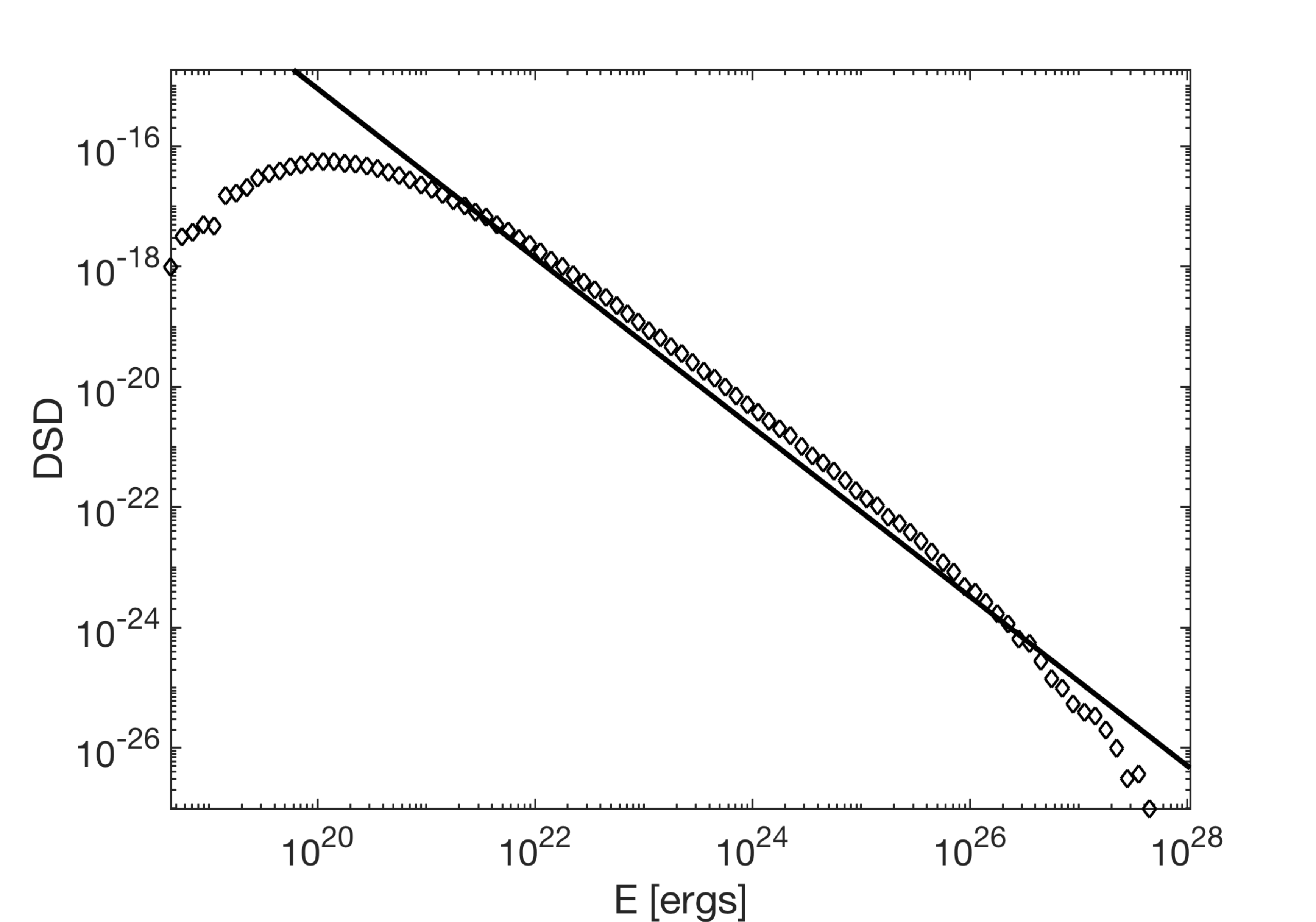}
\caption{Plot of differential size distribution  of the identified features' energy  in logarithmic scale along with the fitted power-law. \label{fig:hist_energy}}
\end{figure}

\begin{figure}
\centering
\includegraphics[width=\hsize]{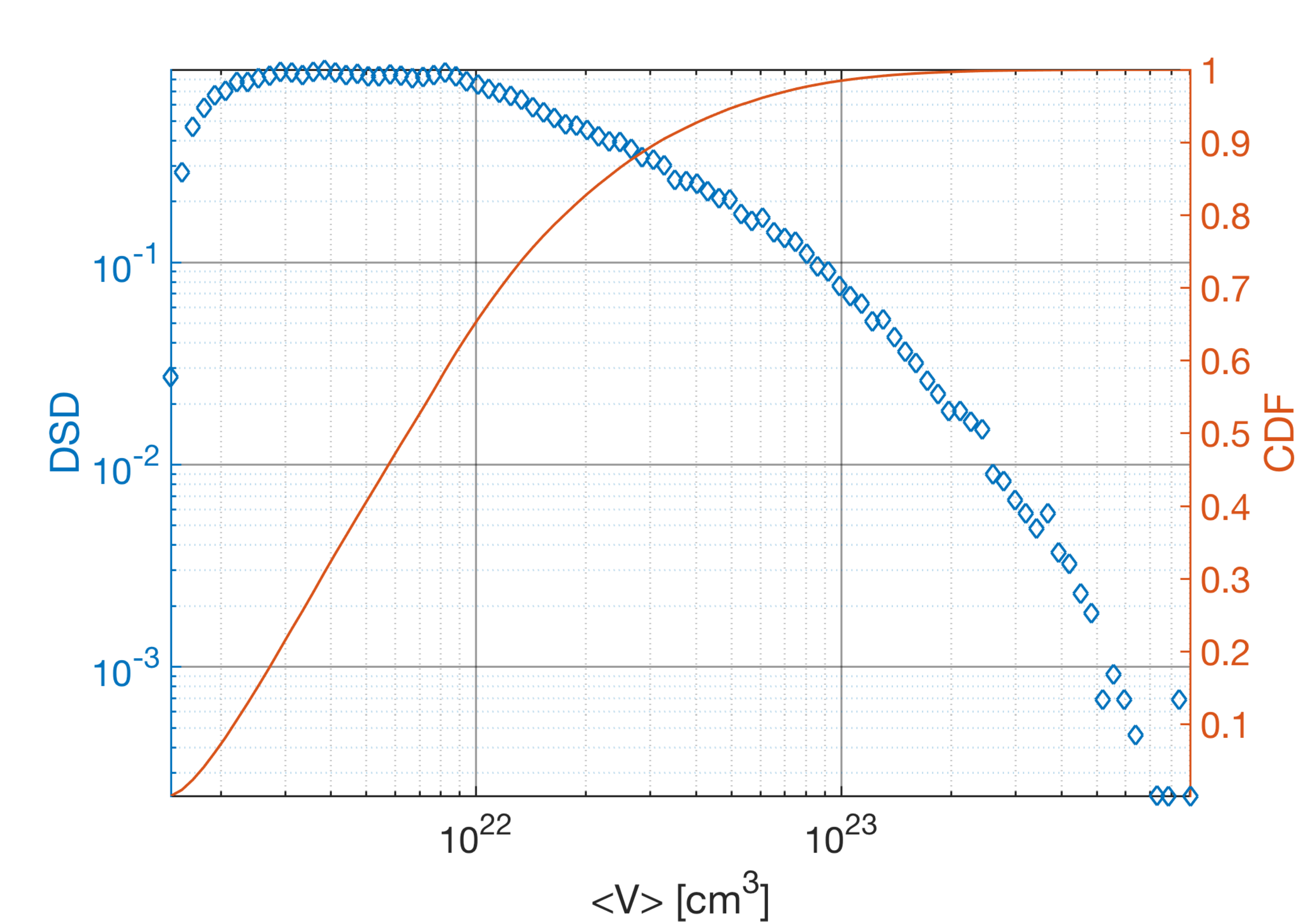}
\caption{Plot of differential size distribution (DSD) and cumulative density function (CDF) of  the averaged volume. DSD is represented with the  left vertical axis, and CDF with the right vertical axis.\label{fig:hist_mean_vol}}
\end{figure}

\begin{figure}
\centering
\includegraphics[width=\hsize]{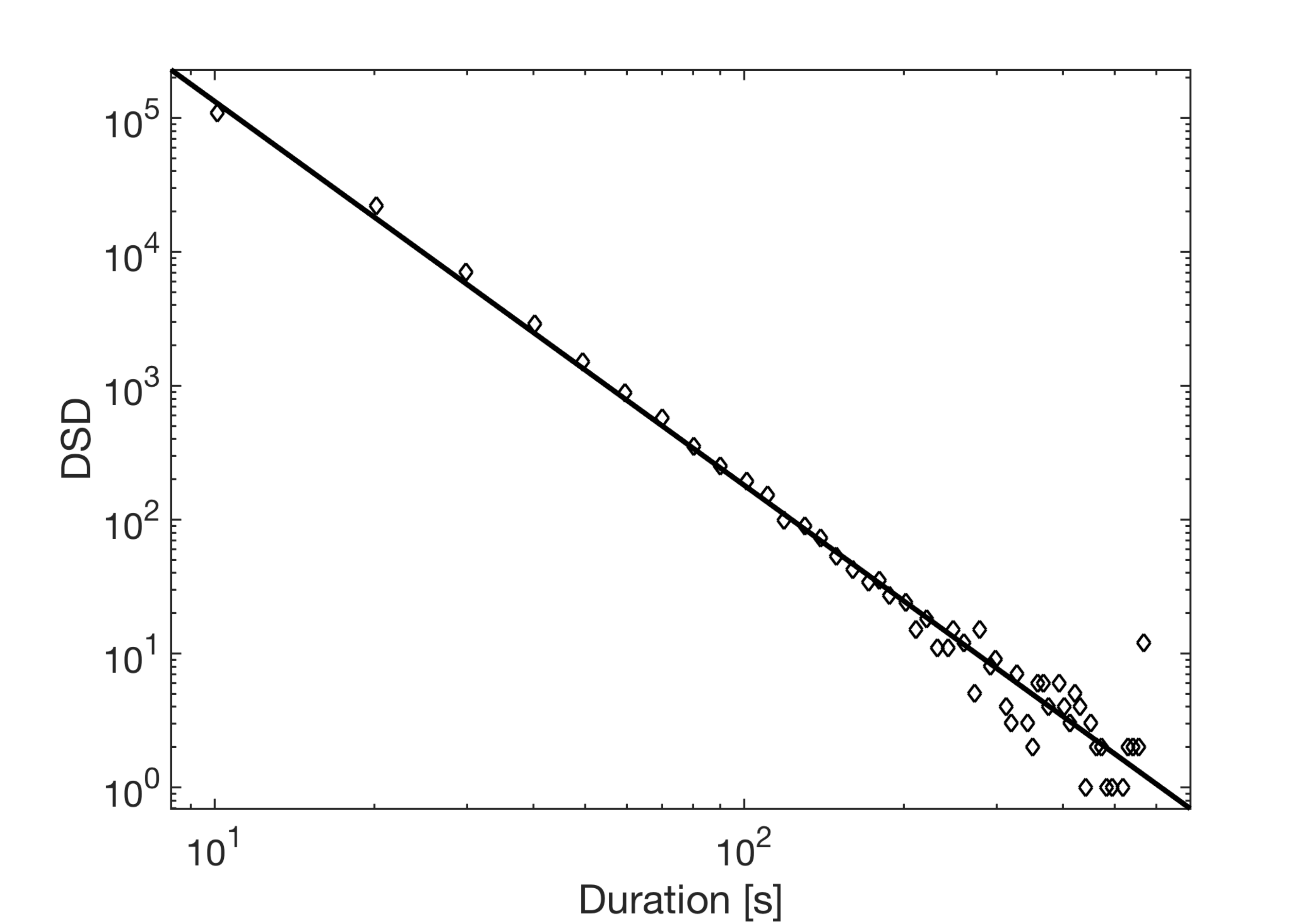}
\caption{Plot of differential size distribution  of events' duration together with a fitted power-law. \label{fig:hist_duration}}
\end{figure}

Power-laws and their indices is a useful tool for the distribution of a quantity, and for checking the importance of smaller scales with respect to larger ones. However, fitting a power-law is sometimes not trivial, and the process usually adds bias to the analysis because it depends on several factors. For example, how well the data are distributed, and what bin-size  and fitting techniques you use. 
Panels a, and b in Fig.  \ref{fig:power_vs_duration_subplot}, show the energy rate evolution of two identified events, but could be a combination of several events occurring successively in close proximity. This happens because the decaying phase of one event overlaps the impulsive phase of another. Our method is not able to resolve the events and they appear as a single event. Being unable to resolve every single event will affect the derived power-laws of all the heating event quantities, such as duration,  energy and volume. The effect on the power-law index can either preserve the index, if small events are just merged into larger events, but does so evenly along the whole energy spectrum, but generally, this induced bias, flattens the powerlaws, but this means that we calculate the lower limit of the power-law indices. 





For general interest, we also look at the events tabulated in the classical event sizes. In table \ref{tbl:tabulated_par_statistics} we have divided the events into three classes: Pico is for events releasing energy less than $10^{24}$ erg, nano for energy release ranging between $10^{24}-10^{27}$ erg, and micro for events spanning between $10^{27}-10^{30}$ erg. We calculate the standard deviation of the duration, the average and total energy and energy rate for each of the classes. We find that $93.5 \ \%$ of the identified events  corresponds to very small events (pico-events) and has an averaged duration equal to 13 s, while nano- ($6.4 \ \%$) and micro-events ($0.03 \ \%$) correspond  to  have averaged durations equal to 48 s and 283 s respectively. The nano events are responsible for releasing most of the energy, followed by the micro flares.

\begin{table}
\begin{center}
\caption{Five  parameters (fraction of events, total, mean ($\mu$), standard deviation ($\sigma$), minimum and maximum value) that describe the three classes of 145306 heating events for duration, rate  of released energy, and released energy. \label{tbl:tabulated_par_statistics}}
\begin{tabular}{cccc}
\hline \hline
&  Pico  & Nano &  Micro    \\
 \hline
 Fract. of events & $93.5 \ \%$ & $6.4 \ \%$  & $0.03 \ \%$ \\ 
 \hline
 & \multicolumn{3}{c}{\textbf{Duration [s]}}\\
\hline

$\mu$              & 13.13    & 48.03 & 283  \\
$\sigma$         & 7.62      & 51.84 & 186 \\
\hline
& \multicolumn{3}{c}{\textbf{Energy Rate [erg/s]}}\\
\hline
Total    &4.27E+26 &4.41E+27 &5.24E+26  \\
$\mu$  &3.14E+21 &4.71E+23 &1.31E+25  \\
\hline
& \multicolumn{3}{c}{\textbf{Energy [erg]}}\\
\hline

Total   &6.75E+27 &2.11E+29 &7.27E+28 \\
$\mu$ &4.97E+22 &2.25E+25 &1.82E+27  \\
\hline
\end{tabular}
\end{center}
\end{table}

\section{Statistical analysis}\label{subsec:statistical_analysis}

The heating events identified can be viewed in to ways. There are the global view, with parameters describing the collection of events, and the local view where the events themselves are analysed. 


To investigate the global view, fig. \ref{fig:timeseries} shows the identified number of features (NOF), the total energy density rate ($Pd_{tot}$), the resolved energy density rate ($Pd_{r}$) and the total volume of the resolved events ($V_{r}$). These parameters are plotted as a function of time, and it can be seen that all of them behaves somewhat stochastically. It can be seen that in broad terms, the number of identified events and their total volume follow each other well, which must mean that the volume distribution is almost constant in time. At the same time, the fraction of the energy density that is identified as events is then also almost constant in time. The combination of the two sets of curves shows that even though the volume of the events are almost constant, both the energy density released and the fraction of that which is identified changes by almost a factor 10. 

The local view compares parameters for each of the identified events. Fig. \ref{fig:dur_vs_en_vs_meanvol}, compares the duration of each of the events with the total energy density of the events, and the average volume of the event. It is interesting to see how large the spread in energy is for the short lived events, where the spread is 7 orders of magnitude, while the longest living events only vary in total energy output by roughly a factor 10. Similarly, the average volume of the events vary by more than two orders of magnitude for the short lived events, while the long lived events are generally all of a volume close to $10^{23}$ cm$^3$. Comparing the average volume with the energy density released by the events, shows again large spreads, but the spread is almost the same for both variables.  


  


\begin{figure}
\centering
\includegraphics[width=\hsize]{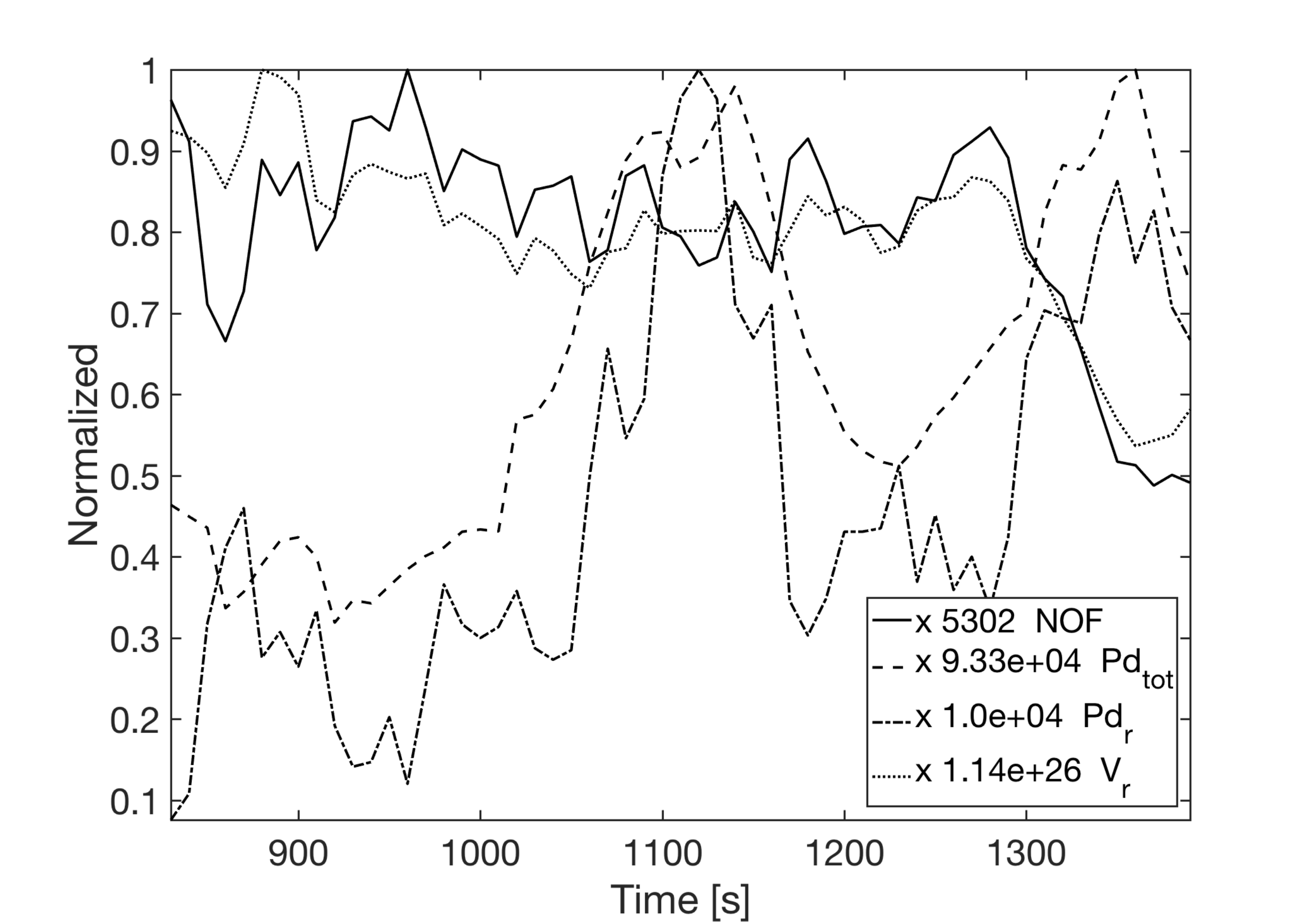}
\caption{Time-series of the following normalised quantities: number of features (full), total energy rate density (dashed), resolved energy density rate (dash-dotted), and resolved volume (dotted). \label{fig:timeseries}}
\end{figure}




\begin{figure}
\begin{subfigure}{0.5\textwidth}
\centering
\includegraphics[width=\hsize]{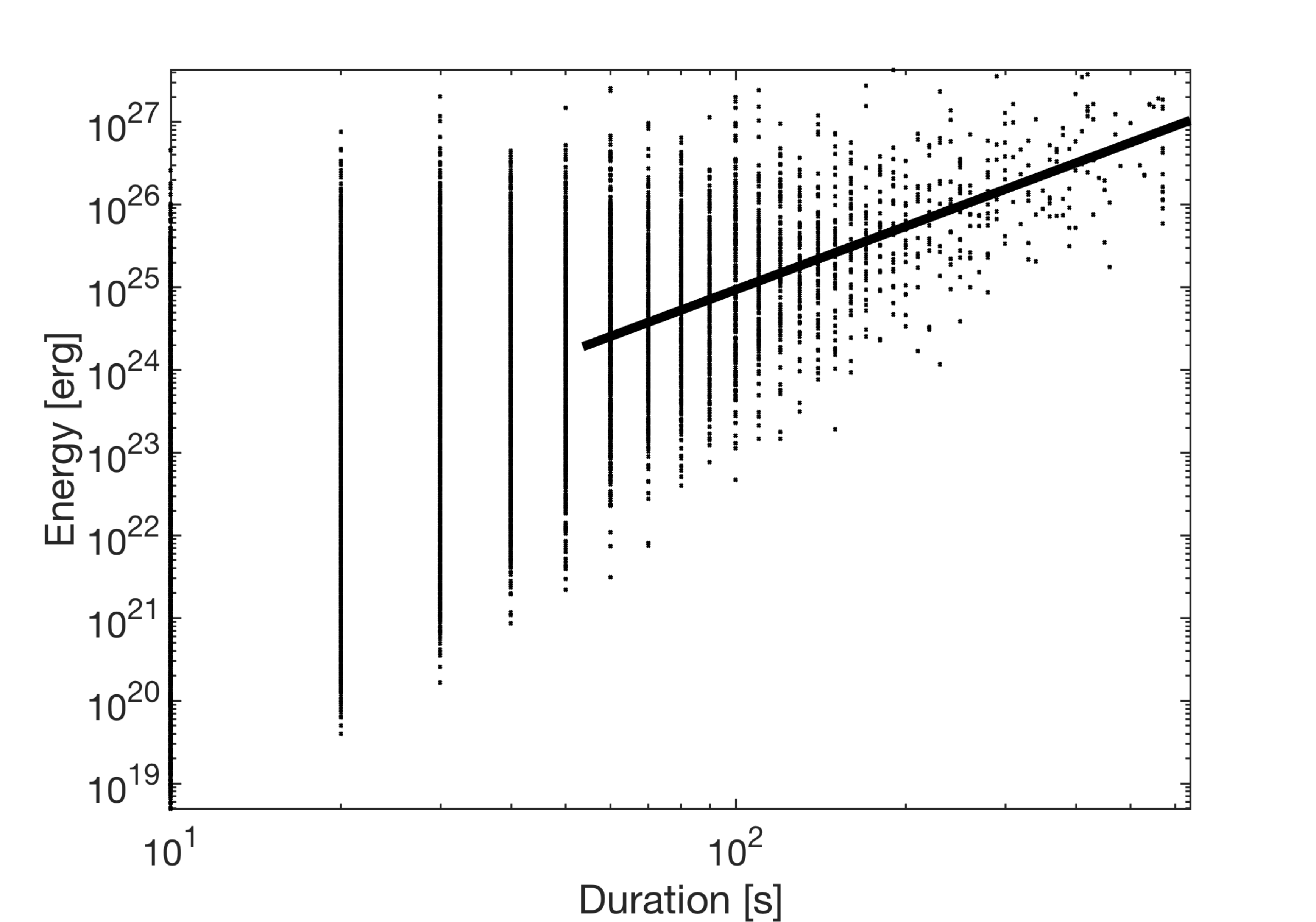}
 \vspace{0.05cm}
\end{subfigure}
\begin{subfigure}{0.5\textwidth}
\centering
\includegraphics[width=\hsize]{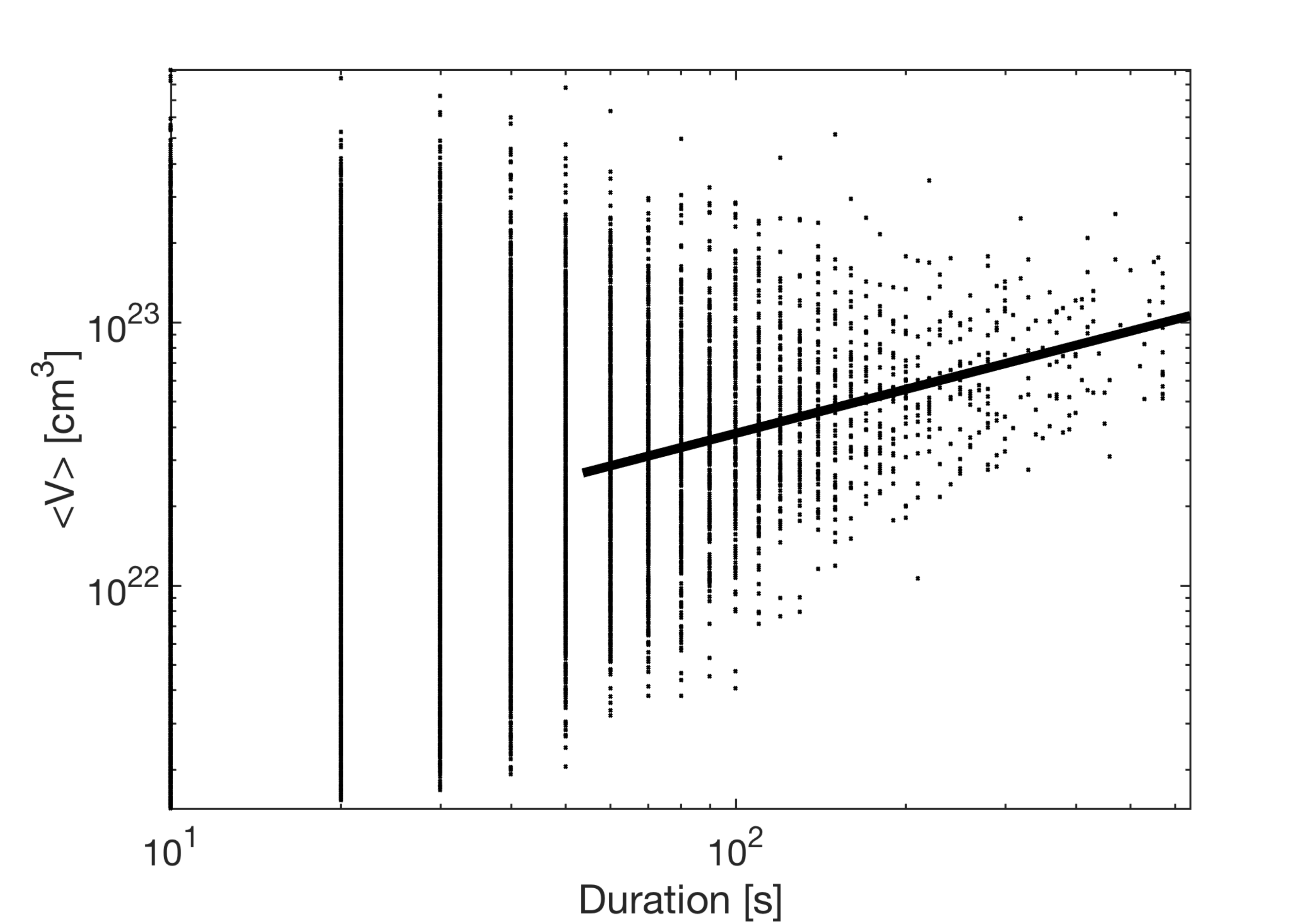}
\end{subfigure}
\caption{Top: Total energy versus duration of all identified features.  A power-law fit is attempted to the data-points that have durations more than 50 s. Those data-points correspond to $0.2 \ \%$ of the total number of events (3083 out of 145306). The power-law index is $\alpha=2.55 \ \pm 0.05$. Bottom: Same as in top panel for mean volume versus duration. The power-law index is $\alpha=0.56 \ \pm 0.01$. \label{fig:dur_vs_en_vs_meanvol}}
\end{figure}

\begin{figure}
\centering
\includegraphics[width=\hsize]{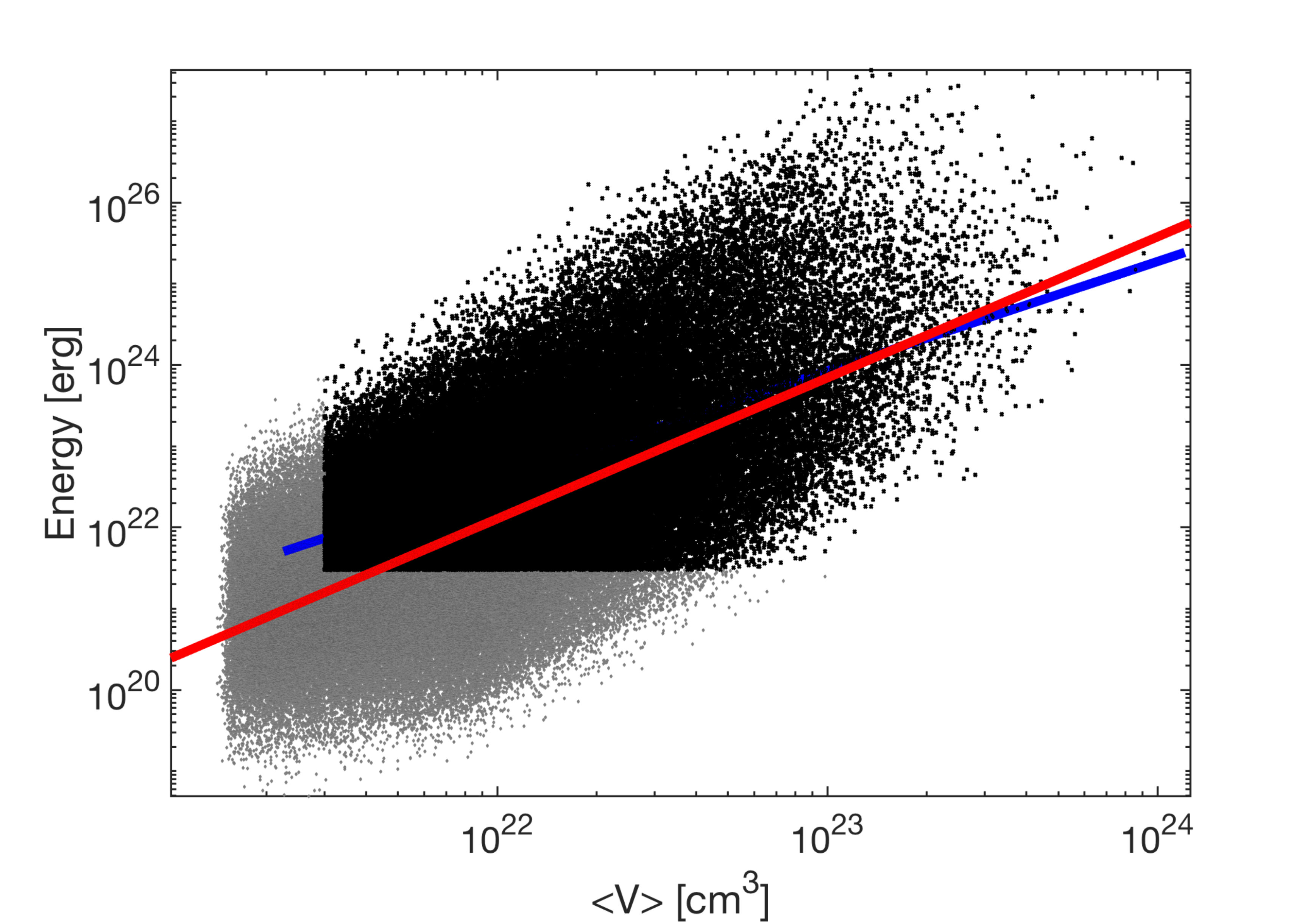}
\caption{Plot of energy versus mean volume together with  power-law fit  of the data-points that correspond to a duration larger than 50 seconds as identified (blue) in Fig. \ref{fig:dur_vs_en_vs_meanvol} and a power-law fit assuming all data-points (red line). The power-law index of the former is $\alpha=1.35 \ \pm 0.01$, and corresponds to $53 \ \%$ of the total number of heating events (76538 out of 145306), while the power-law of the latter is $\alpha=1.74 \ \pm 0.01$  \label{fig:en_vs_meanvol}}
\end{figure}


\section{Discussion and conclusions}\label{sec:conclusions}

The location of Joule heating events is  associated with the magnetic field configuration. It is required that  there is large gradients in the magnetic field, and the magnitude of the event is also dependent on the available magnetic energy. The magnetic energy in the corona is generally a function of the height above the photosphere, and does not vary significantly in the horizontal direction, because the magnetic field dominates the plasma and is configured in a force free state, or at least very close to a force free state \citep{Gudiksen2005}. The gradients in the magnetic field induces currents, which are partly dissipated through electric resistivity. 

In MHD simulations, images such as Figs.  \ref{fig:qj_vs_magnetic_field}  and  \ref{fig:labels_vs_magnetic_field} can shed some light on the details. The general trend is that the most elongated and also largest heating events are formed where the vertical component of the magnetic field, $B_z$, is small and usually at the interfaces between regions with different connectivity (white and faint shades of purple and green areas). Large concentrations of smaller events are present predominately in regions with high magnetic field strength, however the number of events seems to be quenched in regions with the highest flux densities. The explanation for that can be the fact that the stronger the magnetic field, the more difficult is to form magnetic field gradients. 


In this work we have tried to identify as many heating events as possible, using an unbiased method. The events we identify are believed to be mostly reconnection sites, and the reconnection site itself most likely leads to non-uniform heating all the way to the resistive scale. That is caused by the current-sheets being inherently unstable in 3D, creating plasmoids of all sizes in the current sheets \citep{Dahlburg2016}. It is therefore questionable, if we can actually define and identify single events, and that raises the question if the size we attribute to an event is not just a question of resolution.

Other methods which could assign more of the total released Joule energy is possible, but because of our motivation, this method seems the most appropriate. It is extremely difficult to correctly distribute the dissipated energy between the events, making it necessary to discard a large amount of energy in these high dissipation areas. The method we employ is selected to be conservative in the sense that we do not want to mistakenly attribute more energy to an event than we can be certain is part of that single event, and we are able to set strict rules that defines an event. Therefore, the results presented here are unable to give a clear answer to the question of coronal heating being dominated by reconnection events and their distribution. 

The non-constant nature of the identified structures is confirmed in the 2D geometrical parameters of the events cuts in Fig. \ref{fig:geometrical_param}. While identified structures tilt and extend to any direction in the simulation box, the parameters change significantly from one height to another. Such irregularities might occur due to large gradients of the magnetic field, viz. region exhibiting high probability for magnetic reconnection, but could also be evidence for the cross sections of the currents being fractal-like in structure. Another piece of evidence that confirms the clustering of small heating events being very hard to identify as one or several structures, is the multiple peaks in the evolution of energy rate(panels a and b in Fig. \ref{fig:power_vs_duration_subplot}).

The close correlation between the global parameters in Fig. \ref{fig:timeseries} shows that the volume taken up by the heating events and the total number of them, is almost constant in time. In principle, we cannot conclude anything about the distribution of the event volumes from only this evidence but since neither the total volume or the total number of events change, as well as inspection of the differential size distribution for energy shows no difference in shape throughout the simulated timespan we must conclude that the size distribution and energy distribution of the events are both constant, in spite of the large changes in the total energy and resolved energy at about $t = 1100\ $s.

The resolved energy density rate follows the total energy density rate, which can also be seen in Fig. \ref{fig:resolved_E_percentage}. As the method consistently catches roughly 10\% of the released energy, there is a reason to believe that the residual heating is not a due to a  different physical mechanism, as that most likely would not produce a constant ratio when the total energy dissipated by more than a factor two.  

Figs. \ref{fig:dur_vs_en_vs_meanvol} and \ref{fig:en_vs_meanvol} shows that the the energy in the heating events are not given. The total energy delivered by a single heating event is highly dependent on the duration and less dependent on the average volume. Since there is an enormous spread in total energy for heating events of the same duration, it means that the scaling laws between duration, volume are somewhat curious. Initially we imagined that this might be due to the fact that these scaling laws where between integrated values in the 4D space-time, but the scalings between the 4D variables themselves are worse. Trying to produce simple arguments why the scaling law indices have the found magnitudes is not easy, and requires the energy density rates to be complicated functions. Weather these powerlaws are simply an expression of the crowding of many small heating events (Parker's nano-flares) is outside the scope of this paper.


Identifying single or groups of events might affect the power-law distribution of different parameters, such as duration and energy, however we still can derive some conclusions on the impact of heating events on coronal heating. In our results, we observe that the total Joule energy density is smaller by more than two orders of magnitude  than  the  energy density of the magnetic field in the corona  (Fig. \ref{fig:resolved_E_percentage}). As a consequence, only a fraction of the magnetic energy, is needed to heat the corona. A fraction of the total Joule heating in the corona is  attributed to energy released from impulsive events. This fraction varies between 2$ \ \%$ and 14$ \ \%$ indicating the dynamic and intermittent character of heating from impulsive events. In general, the  energy rate related to heating events corresponds to  $8 \ \%$ of the total energy rate of Joule heating released in the corona throughout the total time of investigation. The energy rate released from heating events is approximately $5.4 \ 10^{27}$ erg/s in a volume equal to $24 \times 24 \times 9.5 \ \textrm{Mm}^3$; the resolved energy rate corresponds to energy flux that is $9.4 \  10^{8} \   \textrm{erg cm}^{-2}\textrm{s}^{-1}$. Therefore, the energy flux from impulsive events is two orders of magnitude larger than the typical radiative loss from the quiet Sun, i.e., $ 8 \ 10^5-10^6 \  \textrm{erg cm}^{-2}\textrm{s}^{-1}$ \citep{Withbroe1977,Withbroe1988}. Note however that a big part of that flux is also transported via thermal conduction into the transition region, e.g., pulses of  thermal conduction as described in the dissipative thermal flare model \citep{Brown1979,Smith1979}. 

In this work, we are able to push the lower boundary of identified events down to the energy magnitude of $10^{20}$ erg, i.e., minimum value of \emph{pico}-size events. In addition, we derive duration power index (2.87), which together with and the fact that  $75 \  \%$ of the identified events is not resolved temporally -- they live less than 10\ s -- suggest that the majority of events are short-lived. If this is the case, then observers would need a very short temporal cadence in their observations to capture such short-lived events. Moreover, we find that  our volume data do not follow a power-law distribution, but the cumulative plot  suggests that the majority of events have relatively small volumes. Generally, we find that there is no general rule for how energy is released in individual heating events because results are biased due to event overlapping. In Fig. \ref{fig:power_vs_duration_subplot} for example, we see that small-scale events can be short and impulsive  with single peaks, and their impulsive phase sometimes  last longer than the decay phase, while in some other instances the opposite occurs. These behaviours however, could also be artifacts of the identification method.

Identifying the contribution of small-scale events in heating the solar corona by employing numerical simulations and a conservative identification method has been proven not an easy task.  Certainly, the results are not conclusive, but they point in a certain direction.  Numerous, short-lived, with small spatial extent and stochastic nature is the most abundant population of events in this work. We calculate the energy flux corresponding to nano-events, events with energy within the nanoflare energy range, and we find  that this is more than enough to sustain the energy requirements of the corona. Like observers, we  also identify flat energy power-law distribution. The reason is that small events cluster together forming larger ones. Therefore, an identification method is not able to resolve events temporally and spatially below certain limits due to physical (e.g., background heating) and technical limitations (e.g., threshold criteria). Regardless of the identified sizes of the heating, most of the events occur in regions with low magnetic field, because  magnetic field there can be contorted with ease. However, in regions where the  magnetic field magnitude is large and distortions are harder, the resulting heating events release larger amounts of energy. 

The present work is a first step towards finding the contribution of small scale events related to highly distorted magnetic field in a specific coronal environment, and the values we report seem to be lower limits. It is important for all future investigations of small scale heating events, that the observational and methodological biases are investigated, when an attempt is made to find the elusive power-law index $\alpha$ for the distribution of heating events in the solar corona.  


\begin{acknowledgements} 
This research was supported by the Research Council of Norway through its Centres of Excellence scheme, project number 262622, and through grants of computing time from the Programme for Supercomputing. 
\end{acknowledgements}

\bibliographystyle{aa}
\bibliography{2ndpaper}{}

\begin{thebibliography}{44}
\expandafter\ifx\csname natexlab\endcsname\relax\def\natexlab#1{#1}\fi

\bibitem[{{Aschwanden} {et~al.}(2014){Aschwanden}, {Crosby}, {Dimitropoulou},
  {Georgoulis}, {Hergarten}, {McAteer}, {Milovanov}, {Mineshige}, {Morales},
  {Nishizuka}, {Pruessner}, {Sanchez}, {Sharma}, {Strugarek}, \&
  {Uritsky}}]{Aschwanden2014}
{Aschwanden}, M.~J., {Crosby}, N.~B., {Dimitropoulou}, M., {et~al.} 2014, \ssr
  [\eprint[arXiv]{1403.6528}]

\bibitem[{{Aschwanden} \& {Freeland}(2012)}]{Aschwanden2012}
{Aschwanden}, M.~J. \& {Freeland}, S.~L. 2012, \apj, 754, 112

\bibitem[{{Aschwanden} \& {Parnell}(2002)}]{Aschwanden2002}
{Aschwanden}, M.~J. \& {Parnell}, C.~E. 2002, \apj, 572, 1048

\bibitem[{{Aschwanden} \& {Shimizu}(2013)}]{Shimizu2013}
{Aschwanden}, M.~J. \& {Shimizu}, T. 2013, \apj, 776, 132

\bibitem[{Asgari-Targhi \& van Ballegooijen(2012)}]{AsgariTarghi2012}
Asgari-Targhi, M. \& van Ballegooijen, A.~A. 2012, The Astrophysical Journal,
  746, 81

\bibitem[{{Aulanier} {et~al.}(2006){Aulanier}, {Pariat}, {D{\'e}moulin}, \&
  {DeVore}}]{Aulanier2006}
{Aulanier}, G., {Pariat}, E., {D{\'e}moulin}, P., \& {DeVore}, C.~R. 2006,
  \solphys, 238, 347

\bibitem[{{Benz} \& {Krucker}(2002)}]{Benz2002}
{Benz}, A.~O. \& {Krucker}, S. 2002, \apj, 568, 413

\bibitem[{{Brown} {et~al.}(1979){Brown}, {Spicer}, \& {Melrose}}]{Brown1979}
{Brown}, J.~C., {Spicer}, D.~S., \& {Melrose}, D.~B. 1979, \apj, 228, 592

\bibitem[{{Carlsson} {et~al.}(2007){Carlsson}, {Hansteen}, {de Pontieu},
  {McIntosh}, {Tarbell}, {Shine}, {Tsuneta}, {Katsukawa}, {Ichimoto},
  {Suematsu}, {Shimizu}, \& {Nagata}}]{Carlsson2007}
{Carlsson}, M., {Hansteen}, V.~H., {de Pontieu}, B., {et~al.} 2007, \pasj, 59,
  S663

\bibitem[{{Carlsson} {et~al.}(2016){Carlsson}, {Hansteen}, {Gudiksen},
  {Leenaarts}, \& {De Pontieu}}]{Carlsson2016}
{Carlsson}, M., {Hansteen}, V.~H., {Gudiksen}, B.~V., {Leenaarts}, J., \& {De
  Pontieu}, B. 2016, \aap, 585, A4

\bibitem[{Carlsson \& Stein(2002)}]{Carlsson2002}
Carlsson, M. \& Stein, R. 2002, The Astrophysical Journal, 572, 626

\bibitem[{{Charalambos} \& {Gudiksen}(2017)}]{Kanella2017a}
{Charalambos}, K. \& {Gudiksen}, B. 2017, Astron. Astrophys., 603, A83

\bibitem[{{Christe} {et~al.}(2008){Christe}, {Hannah}, {Krucker}, {McTiernan},
  \& {Lin}}]{Christe2008}
{Christe}, S., {Hannah}, I.~G., {Krucker}, S., {McTiernan}, J., \& {Lin}, R.~P.
  2008, \apj, 677, 1385

\bibitem[{Collins(2007)}]{Collins2007}
Collins, T.~J. 2007, Biotechniques, 43, 25

\bibitem[{Crosby {et~al.}(1993)Crosby, Aschwanden, \& Dennis}]{Crosby1993}
Crosby, N.~B., Aschwanden, M.~J., \& Dennis, B.~R. 1993, Solar Physics, 143,
  275

\bibitem[{Dahlburg {et~al.}(2016)Dahlburg, Einaudi, Taylor, Ugarte-Urra,
  Warren, Rappazzo, \& Velli}]{Dahlburg2016}
Dahlburg, R.~B., Einaudi, G., Taylor, B.~D., {et~al.} 2016, Astrophys. J., 817,
  47

\bibitem[{{Drake}(1971)}]{Drake1971}
{Drake}, J.~F. 1971, \solphys, 16, 152

\bibitem[{{Galsgaard} \& {Nordlund}(1996)}]{Galsgaard1996}
{Galsgaard}, K. \& {Nordlund}, {\AA}. 1996, \jgr, 101, 13445

\bibitem[{{Gesztelyi} {et~al.}(1986){Gesztelyi}, {Gerlei}, {Karlicky},
  {Farnik}, \& {Valnicek}}]{Gesztelyi1986}
{Gesztelyi}, L., {Gerlei}, O., {Karlicky}, M., {Farnik}, F., \& {Valnicek}, B.
  1986, in The lower atmosphere of solar flares; Proceedings of the Solar
  Maximum Mission Symposium, Sunspot, NM, Aug. 20-24, 1985 (A87-26201 10-92).
  Sunspot, NM, National Solar Observatory, 1986, p. 163-177., ed. D.~F.
  {Neidig}, 163--177

\bibitem[{{Gudiksen} {et~al.}(2011){Gudiksen}, {Carlsson}, {Hansteen}, {Hayek},
  {Leenaarts}, \& {Mart{\'{\i}}nez-Sykora}}]{Gudiksen2011}
{Gudiksen}, B.~V., {Carlsson}, M., {Hansteen}, V.~H., {et~al.} 2011, \aap, 531,
  A154

\bibitem[{{Gudiksen} \& {Nordlund}(2005)}]{Gudiksen2005}
{Gudiksen}, B.~V. \& {Nordlund}, {\AA}. 2005, Apj, 618, 1020

\bibitem[{Gul-Mohammed {et~al.}(2014)Gul-Mohammed, Arganda-Carreras, Andrey,
  Galy, \& Boudier}]{Gul-Mohammed2014}
Gul-Mohammed, J., Arganda-Carreras, I., Andrey, P., Galy, V., \& Boudier, T.
  2014, BMC Bioinformatics, 15, 9

\bibitem[{{Hannah} {et~al.}(2011){Hannah}, {Hudson}, {Battaglia}, {Christe},
  {Ka{\v s}parov{\'a}}, {Krucker}, {Kundu}, \& {Veronig}}]{Hannah2011}
{Hannah}, I.~G., {Hudson}, H.~S., {Battaglia}, M., {et~al.} 2011, \ssr, 159,
  263

\bibitem[{{Hansteen} {et~al.}(2015){Hansteen}, {Guerreiro}, {De Pontieu}, \&
  {Carlsson}}]{Hansteen2015}
{Hansteen}, V., {Guerreiro}, N., {De Pontieu}, B., \& {Carlsson}, M. 2015,
  \apj, 811, 106

\bibitem[{{Hara} \& {Ichimoto}(1999)}]{Hara1999}
{Hara}, H. \& {Ichimoto}, K. 1999, \apj, 513, 969

\bibitem[{{Hayek} {et~al.}(2010){Hayek}, {Asplund}, {Carlsson}, {Trampedach},
  {Collet}, {Gudiksen}, {Hansteen}, \& {Leenaarts}}]{Hayek2010}
{Hayek}, W., {Asplund}, M., {Carlsson}, M., {et~al.} 2010, \aap, 517, A49

\bibitem[{{Hudson}(1991)}]{Hudson1991}
{Hudson}, H.~S. 1991, \solphys, 133, 357

\bibitem[{{Janvier} {et~al.}(2014){Janvier}, {Aulanier}, {Bommier},
  {Schmieder}, {D{\'e}moulin}, \& {Pariat}}]{Janvier2014}
{Janvier}, M., {Aulanier}, G., {Bommier}, V., {et~al.} 2014, \apj, 788, 60

\bibitem[{Klimchuk(2006)}]{Klimchuk2006}
Klimchuk, J.~A. 2006, Solar Physics, 234, 41

\bibitem[{{Low}(1990)}]{Low1990}
{Low}, B.~C. 1990, \araa, 28, 491

\bibitem[{{Morales} \& {Charbonneau}(2009)}]{Morales2009}
{Morales}, L. \& {Charbonneau}, P. 2009, \apj, 698, 1893

\bibitem[{{Nordlund} \& {Galsgaard}(2012)}]{Nordlund2012}
{Nordlund}, A. \& {Galsgaard}, K. 2012, in EGU General Assembly Conference
  Abstracts, Vol.~14, EGU General Assembly Conference Abstracts, ed.
  A.~{Abbasi} \& N.~{Giesen}, 12646

\bibitem[{Ollion {et~al.}(2013)Ollion, Cochennec, Loll, Escud{\'{e}}, \&
  Boudier}]{Ollion2013}
Ollion, J., Cochennec, J., Loll, F., Escud{\'{e}}, C., \& Boudier, T. 2013,
  Bioinformatics (Oxford, England), 29, 1840

\bibitem[{Parker(1972)}]{Parker1972}
Parker, E.~N. 1972, Astrophysical Journal, 174, 499

\bibitem[{Parker(1983)}]{Parker1983b}
Parker, E.~N. 1983, Astrophysical Journal, 264

\bibitem[{{Parker}(1988)}]{Parker1988}
{Parker}, E.~N. 1988, \apj, 330, 474

\bibitem[{{P{\'e}rez Enriquez} \& {Miroshnichenko}(1999)}]{Enriquez1999}
{P{\'e}rez Enriquez}, R. \& {Miroshnichenko}, L.~I. 1999, \solphys, 188, 169

\bibitem[{{Smith} \& {Lilliequist}(1979)}]{Smith1979}
{Smith}, D.~F. \& {Lilliequist}, C.~G. 1979, \apj, 232, 582

\bibitem[{{Spitzer}(1962)}]{Spitzerbook}
{Spitzer}, L. 1962, {Physics of Fully Ionized Gases}

\bibitem[{{Tomczyk} {et~al.}(2007){Tomczyk}, {McIntosh}, {Keil}, {Judge},
  {Schad}, {Seeley}, \& {Edmondson}}]{Tomczyk2007}
{Tomczyk}, S., {McIntosh}, S.~W., {Keil}, S.~L., {et~al.} 2007, Science, 317,
  1192

\bibitem[{{van Ballegooijen}(1986)}]{VanBallegooijen1986}
{van Ballegooijen}, A.~A. 1986, Astrophysical Journal, 311, 1001

\bibitem[{van Ballegooijen {et~al.}(2011)van Ballegooijen, Asgari-Targhi,
  Cranmer, \& DeLuca}]{vanBallegooijen2011}
van Ballegooijen, A.~A., Asgari-Targhi, M., Cranmer, S.~R., \& DeLuca, E.~E.
  2011, The Astrophysical Journal, 736, 3

\bibitem[{{Withbroe}(1988)}]{Withbroe1988}
{Withbroe}, G.~L. 1988, \apj, 325, 442

\bibitem[{Withbroe \& Noyes(1977)}]{Withbroe1977}
Withbroe, G.~L. \& Noyes, R.~W. 1977, Annual Review of Astronomy and
  Astrophysics, 15, 363

\end{thebibliography}

\end{document}